\documentclass[a4paper,11pt]{article}
 \pdfoutput=1
\usepackage{amssymb}
\usepackage{pseudocode}
\usepackage{graphicx}
\usepackage[caption=false]{subfig}
\usepackage{multirow}
\usepackage{url}

\begin{document}


\renewcommand{\thefootnote}{\fnsymbol{footnote}}
\title{Automatic Construction of Statechart-Based \\ Anomaly Detection Models for \\ Multi-Threaded Industrial Control Systems\footnotemark[1]}

\author{Amit Kleinmann, Avishai Wool\\
       Tel-Aviv University, Tel-Aviv 69978, Israel\\
       \texttt{amitkl@post.tau.ac.il},
       \texttt{yash@eng.tau.ac.il}
}

\maketitle

\footnotetext[1]{
Supported in part by a grant from the Israeli Ministry of Science and Technology and by a grant from the Interdisciplinary Cyber Research Center at TAU. Some of the results were presented at the The 10th International Conference on Critical Information Infrastructures Security (CRITIS 2015).
}
\begin{abstract}
Traffic of Industrial Control System (ICS) between the Human Machine Interface (HMI) and the Programmable Logic Controller (PLC) is known to be highly periodic. However, it is sometimes multiplexed, due to asynchronous scheduling. Modeling the network traffic patterns of multiplexed ICS streams using Deterministic Finite Automata (DFA) for anomaly detection typically produces a very large DFA, and a high false-alarm rate. In this paper we introduce a new modeling approach that addresses this gap. Our {\em Statechart DFA} modeling includes multiple DFAs, one per cyclic pattern, together with a DFA-selector that de-multiplexes the incoming traffic into sub-channels and sends them to their respective DFAs. We demonstrate how to automatically construct the statechart from a captured traffic stream. Our unsupervised learning algorithms first build a Discrete-Time Markov Chain (DTMC) from the stream. Next we split the symbols into sets, one per multiplexed cycle, based on symbol frequencies and node degrees in the DTMC graph. Then we create a sub-graph for each cycle, and extract Euler cycles for each sub-graph. The final statechart is comprised of one DFA per Euler cycle. The algorithms allow for non-unique symbols, that appear in more than one cycle, and also for symbols that appear more than once in a cycle.

We evaluated our solution on traces from a production ICS using the Siemens S7-0x72 protocol. We also stress-tested our algorithms on a collection of synthetically-generated traces that simulated multiplexed ICS traces with varying levels of symbol uniqueness and time overlap. The algorithms were able to split the symbols into sets with 99.6\% accuracy. The resulting statechart modeled the traces with a median false-alarm rate of as low as 0.483\%. In all but the most extreme scenarios the {\em Statechart} model drastically reduced both the false-alarm rate and the learned model size in comparison with the naive single-DFA model.
\end{abstract}

\providecommand{\keywords}[1]{\textbf{\textit{Index terms---}} #1}

\section{Introduction}
Industrial Control Systems (ICS) are used for monitoring and controlling numerous industrial systems and processes. In particular, ICS are used in critical infrastructure assets such as chemical plants, electric power generation, transmission and distribution systems, water distribution networks, and waste water treatment facilities. ICS have a strategic significance due to the potentially serious consequences of a fault or malfunction.

\subsection{Background}
ICS typically incorporate sensors and actuators that are controlled by Programmable Logic Controllers (PLCs), and which are themselves managed by a Human Machine Interface (HMI). PLCs are computer-based devices that were originally designed to perform the logic functions executed by electrical hardware (relays, switches, and mechanical timer/counters).  PLCs have evolved into controllers with the capability of controlling the complex processes used for discrete control in discrete manufacturing.  The NIST Guide to ICS Security \cite{stouffer} explains that ICS is a general term that encompasses several types of control systems, including Programmable Logic Controllers (PLC), Distributed Control Systems (DCS), Supervisory Control And Data Acquisition (SCADA) systems, and other control system configurations. An automation system within a campus is usually referred to as a DCS, while SCADA systems typically comprise of different stations distributed over large geographical areas.

ICS were originally designed for serial communications, and were built on the premise that all the operating entities would be legitimate, properly installed, perform the intended logic and follow the protocol. Thus, many ICSes  have almost no measures for defending against deliberate attacks. Specifically, ICS network components do not verify the identity and permissions of other components with which they interact (i.e., no authentication and authorization mechanisms); they do not verify message content and legitimacy (i.e., no data integrity checks); and all the data sent over the network is in plaintext (i.e., no encryption to preserve confidentiality). Therefore, deploying an Intrusion Detection Systems (IDS) in an ICS network is an important defensive measure.

\subsection{Related work}
\cite{byres2004use} describe different attack trees on ICS based on the Modbus/TCP protocol. They found that compromising the slave (PLC) or the master (HMI) has the most severe potential impact on the ICS. For instance, an attacker that gains access to the ICS could identify as the HMI and change data values in the PLC. Alternately, an attacker can perform a Man In The Middle attack between a PLC and HMI and ``feed'' the HMI with misleading data, allegedly coming from the exploited PLC.

While most of the current commercial network intrusion detection systems (NIDS) are signature-based, i.e., they recognize an attack when it matches a previously defined signature, Anomaly-based Network Intrusion Detection Systems (IDS) ``are based on the belief that an intruder's behavior will be noticeably different from that of a legitimate user'' \cite{mukherjee1994network}.

\cite{yang2006anomaly} discussed these two potential types of IDS for ICS, stating that ``signature detection recognizes an intrusion based on known intrusion or attack characteristics or signatures. Anomaly detection identifies an intrusion by calculating a deviation from normal system behavior''.

Different kinds of Anomaly Intrusion Detection models have been suggested for SCADA systems. \cite{yang2006anomaly} used an Auto Associative Kernel Regression (AAKR) model coupled with the Statistical Probability Ratio Test (SPRT) and applied them on a SCADA system looking for matching patterns. The model used numerous indicators representing network traffic and hardware-operating statistics to predict the `normal' behavior.

Several recent studies \cite{atassi,chen} suggest anomaly-based detection for SCADA systems which are based on Markov chains. However, \cite{ye} showed that although the detection accuracy of this technique is high, the number of False Positive values is also high, as it is sensitive to noise. \cite{hadziosmanovic2011} used the logs generated by the control application running on the HMI to detect anomalous patterns of user actions on process control application.

\cite{fovino2010modbus} have presented a state-based intrusion detection system for SCADA systems. Their approach uses detailed knowledge of the industrial process' control to generate a system virtual image. The virtual image represents the PLCs 
of a monitored system, with all their memory registers, coils, inputs and outputs. The virtual image is updated using a periodic active synchronization procedure and via a feed generated by the intrusion detection system (i.e., known intrusion signatures).

Model-based anomaly detection for SCADA systems, and specifically for Modbus traffic, was introduced by ~\cite{cheung2007using}. They designed a multi-algorithm intrusion detection appliance for Modbus/TCP with pattern anomaly recognition, Bayesian analysis of TCP headers and stateful protocol monitoring, complemented with customized Snort rules \cite{roesch}. In subsequent work, \cite{valdes2009communication} incorporated adaptive statistical learning methods into the system to detect for communication patterns among hosts and traffic patterns in individual flows. Later \cite{briesemeister2010detection} integrated these  intrusion detection technologies into the EMERALD event correlation framework \cite{emerald}.

A survey of techniques related to learning and detection of anomalies in critical control systems can be found in \cite{alcaraz_cazorla_fernandez}.

\cite{sommer} discuss the surprising imbalance between the extensive amount of research on machine learning-based anomaly detection pursued in the academic intrusion detection community, versus the lack of operational deployments of such systems. One of the reasons for that, by the authors, is that the machine learning anomaly detection systems are lacking the ability to bypass the ``semantic gap'': The system ``understands'' that an abnormal activity has occurred, but it cannot produce a message that will elaborate, helping the operator differentiate between an abnormal activity and an attack.

\cite{Erez:2015:CVC:2822917.2823033} developed an anomaly detection system that detects irregular changes in SCADA control registers' values. The system is based on an automatic classifier that identifies several classes of PLC registers (Sensor registers, Counter registers and Constant registers). Parameterized behavior models were built for each class. In its learning phase, the system instantiates the model for each register. During the enforcement phase the system detects deviations from the model.

\cite{Goldenberg201363} developed a model-based approach (the GW model) for Network Intrusion Detection based on the normal traffic pattern in Modbus SCADA Networks using a DFA to represent the cyclic traffic. \\
Subsequently, \cite{KleinmannWjdfsl} demonstrated that a similar methodology is successful also in SCADA systems running the Siemens S7 protocol. 

\cite{Caselli} proposed a methodology to model sequences of SCADA protocol messages as Discrete Time Markov Chains (DTMCs). They built a state machine whose states model possible messages, and whose transitions model a ``followed-by'' relation. Based on data from three different Dutch utilities the authors found that only 35\%-75\% of the possible transitions in the DTMC were observed.  This strengthens the observations of \cite{Goldenberg201363,KleinmannWjdfsl} of a substantial sequentiality in the SCADA communications. However, unlike \cite{Goldenberg201363,KleinmannWjdfsl} they did not observe clear cyclic message patterns. The authors hypothesized that the difficulties in finding clear sequences is due to the presence of several threads in the HMI's operating system that multiplex requests on the same TCP stream.

Modeling the network traffic patterns of multiplexed SCADA streams, as observed by \cite{Caselli}, using Deterministic Finite Automata (DFA) for anomaly detection typically produces a very large DFA, and a high false-alarm rate.

\subsection{Contributions}
DFA-based models have been shown to be extremely effective in modeling the network traffic patterns of SCADA systems \cite{Goldenberg201363,KleinmannWool2015}, thus allowing the creation of anomaly-detection systems with low false-alarm rates. However, the existing DFA-based models can be improved in some scenarios.

In this paper we address two such scenarios: the first scenario is the one identified in~\cite{Caselli}: the HMI is multi-threaded, each thread independently scans a separate set of control registers, and each thread has its own scan frequency. The second scenario occurs when the SCADA protocol allows the HMI to ``subscribe'' to a certain register range, after which the PLC asynchronously sends a stream of notifications with the values of the subscribed registers. The commonality between the scenarios is that the network traffic is not the result of a single cyclic pattern: it is the result of several multiplexed cyclic patterns. The multiplexing is due to the asynchronous scheduling of the threads inside the HMI, or to the asynchronous scheduling of PLC-driven notifications. Attempting to model a multiplexed stream by a single DFA typically produces a very large DFA (it's cycle length can be the least-common-multiple of the individual cycle lengths), and also a high false-alarm rate because of the variations in the scheduling of the independent threads.

Our solution to both scenarios is the same: instead of modeling the traffic of an HMI-PLC channel by a single DFA, we model it as a {\em Statechart} \cite{Harel1987} of multiple DFAs, one per cyclic pattern, with a DFA-selector that de-multiplexes the incoming stream of symbols (messages) into sub-channels and sends them to their respective DFAs.

In the most simple cases, the number of cyclic patterns is known and each of the cyclic patterns contains easy-to-identify distinct symbols, i.e., symbols that do not appear in any of the other cyclic patterns. In these cases, each cyclic pattern can be learned individually based on some easy to detect meta data, e.g., the SCADA traffic can be demultiplexed to sub streams, each containing traffic of individual cyclic pattern, and then each of the cyclic pattern can be learned out of its corresponding sub-stream. In more complex cases the same symbols appear in several cyclic pattern, thus some of the patterns overlap. Our design supports both the simple cases, in which each sub-channel has a unique set of symbols---and also the complex cases in which some symbols belong to multiple cyclic patterns. We suggest a new method to identify the number of cycles and to learn each of the multiplexed cyclic patterns even in cases where there is symbol overlaps between different patterns. Our learning algorithms first build a Discrete-Time Markov Chain (DTMC) from the stream. Next we split the symbols into sets, one per multiplexed cycle, based on symbol frequencies and node degrees in the DTMC graph. Then we create a sub-graph for each cycle, and extract Euler cycles for each sub-graph. The final {\em Statechart} is comprised of one DFA per Euler cycle.

We evaluated our solution on traces from a production SCADA system using the latest variant of the proprietary Siemens S7 protocol, so called S7-0x72. Unlike the standard  S7-0x32 protocol, which is fairly well understood, little is published about the new variant. Based on recent advances in the development of an open-source Wireshark dissector for this variant, we were able to model S7-0x72 in the {\em Statechart} framework, including its subscribe/notify capability.  A naive single-DFA model caused a false-alarm rate of 13--14\% on our traces, while the {\em Statechart} model reduced the false-alarm rate by two orders of magnitude, down to at most 0.11\%. A separate contribution is our description of the S7-0x72 protocol, with its complex message formats and advanced semantics.

We also stress--tested our solution on a collection of synthetically--generated traces, with intentionally difficult scenarios multiplexing up to 4 periodic patterns and with up to 56\% symbol overlap between patterns.
Our learning algorithms were able to correctly identify the set of symbols in each cycle in 99.6\% of the cases.
In all but the most extreme scenarios the {\em Statechart} model drastically reduced both the false-alarm rate and the model size in comparison with the naive single-DFA model.
The resulting Statecharts were able to model the multiplexed stream with an average of only 3.3\% more false alarms than the ideal Statechart (in which the true cycles are known) even when the cycles did not have unique symbol sets, and when a symbol appears multiple times in the cycles. The overall median false-alarm rate was as low as 0.483\%.

\section{Preliminaries}
\subsection{The DFA-based model for Modbus}
The GW model \cite{Goldenberg201363} was developed and tested on Modbus traffic. Modbus is a simple request-response protocol widely used in SCADA networks.
A typical SCADA HMI sends a request to a PLC. The request includes a function code specifying the service, and the address range of data items. After the PLC processes the request, it sends a response back to the HMI.

In the GW model, the key assumption is that traffic is {\em periodic}, therefore, each HMI-PLC channel is modeled by a Mealy Deterministic Finite Automaton (DFA). The DFA for Modbus has the following characteristics:

(a) A symbol is defined as a concatenation of the message type, function code, and address range, totaling 33-bits;

(b) A state is defined for each message in the periodic traffic pattern.

The GW model suggests a network anomaly detection system that comprises two stages: An unsupervised learning stage, and an enforcement stage. In the unsupervised learning stage a fixed number of messages is captured, the pattern
length is revealed, and Mealy DFA is built for each HMI-PLC channel. The channel's input-symbols are categorized into two groups: Known and Un-known. The Known group consists of all the input symbols that were observed during the learning phase, and have a matching DFA state. The Unknown symbols are all the rest.

\begin{figure}[t]
\centering
   \includegraphics[ width=0.95\textwidth,natwidth=500,natheight=60,
   trim = 30 160 20 100]{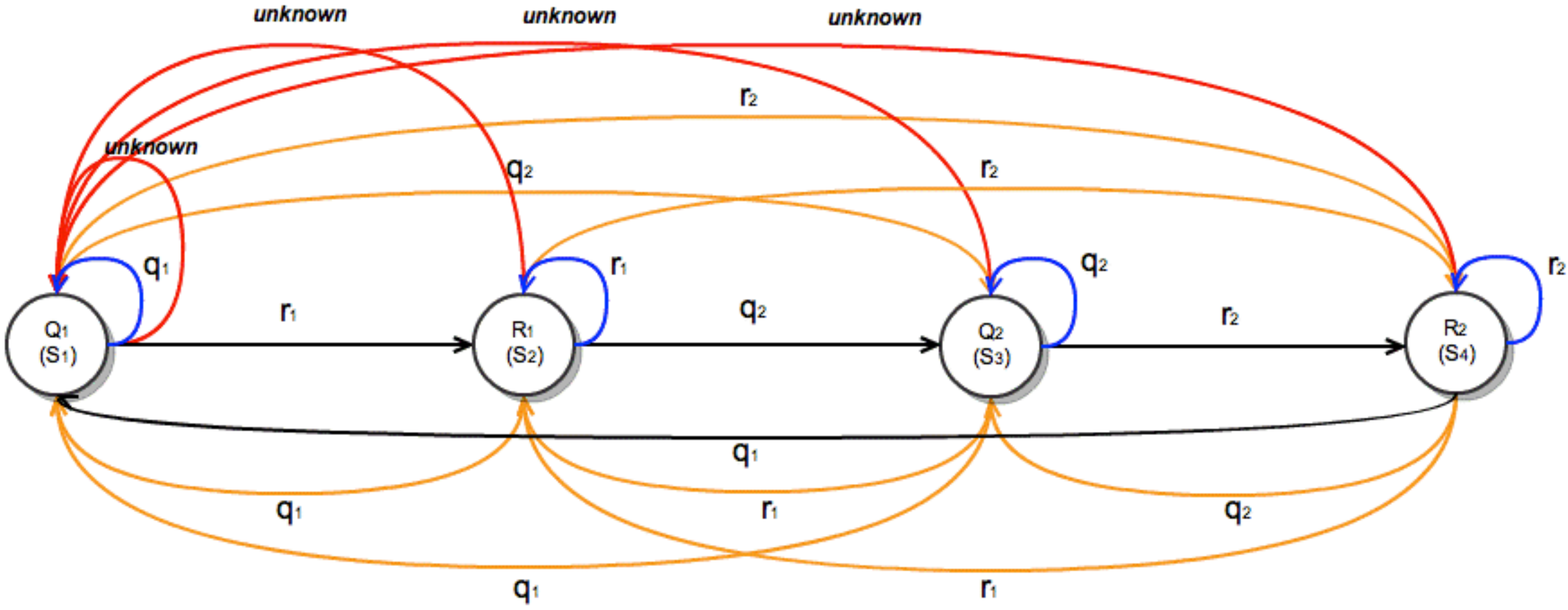}
   \hfill
\caption{A DFA representing a 2-Query SCADA traffic pattern.}
    \label{fig:DFA}
\end{figure}

We denote current position as $S_i$ and the received input symbol as $s_j$. Four transition types (output symbols) are defined in the DFA (as depicted at Figure \ref{fig:DFA}):
\begin{itemize}
\item {\bf Normal} - A ``normal'' transition occures on a known symbol that leads to the next state in the periodic sequence. I.e., $s_j = S_{i+1}$. As a result of a ``normal'' event the DFA is transitioned to its next state $S_{i+1}$.
\item {\bf Retransmission} - A ``retransmission'' is an occurrence of a known symbol that is identical to the previous symbol. I.e., $s_j = S_i$. As a result of a ``retransmission'' event the DFA remains in its current state $S_{i}$. Note that if the pattern includes two identical symbols it would lead to a state with 2 different transitions for the same symbol (a ``normal'' transition forward, and a ``retransmission'' self loop). This non-determinism is resolved in run-time by preferring the ``normal'' transition over the self-loop ``retransmission'' transition.
\item {\bf Miss} - A ``miss'' is an occurrence of a known symbol $s_j$ which appears at state $S_i$ out of its expected position in the pattern. I.e., $s_j \neq S_{i+1}$. As a result of a ``miss'' event the DFA is transitioned to the closest forward state (modulu {\it Pattern\_Length}) that follows the normal $s_j$ symbol.
\item {\bf Unknown} - An ``unknown'' is an occurrence of an unknown symbol. As a result of an ``unknown'' event the DFA remains in its current state $S_{i}$.
\end{itemize}

The learning assumes that the sniffed traffic is benign. In the enforcement stage, traffic is monitored for each channel (according to its DFA), and proper events are triggered.

Based on traffic captured from a production Modbus system, the authors discovered that over 97\% of Modbus traffic is well modeled by a single DFA per HMI-PLC channel. However they also discovered a phenomenon that challenges the DFA-based approach: In addition to a frequent scan cycle that occurs multiple time per second, they found a second periodic pattern with a 15-minute cycle. Attempting to model both cycles by a single DFA produces a very large, unwieldy model: Its normal pattern consists of hundreds of repetitions of the fast scan cycle followed by one repetition  of the slow cycle. Such a pattern is also inaccurate since the slow cycle does not always interrupt the fast cycle at the same point, and while the slow pattern is active, symbols from both patterns are interleaved.

\subsection{Adversary model}
In this work we assume the existence of a semantic adversary who has unrestricted physical access to the SCADA network and has thus nearly complete control of the communications channel between the HMI and the PLCs. Our underlying threat model is based on the Dolev-Yao threat model \cite{Dolev:1981:SPK:891726}. The adversary, may overhear and intercept all traffic regardless of its source and destination. The adversary can inject arbitrary packets with any source and destination addresses. Consequently, the adversary can also replay previously overheard messages. In particular the adversary can take over the HMI and issue control messages. The objective of the adversary is to manipulate the SCADA network to achieve an impact on the physical world.

Currently, most SCADA protocols do not include cryptographic algorithms (e.g., ciphers and hash functions). Our adversary model assumes that when such security measures are deployed, their associated cryptographic keys are known to (or can be broken by) the adversary. However, the adversary is limited by the cryptographic methods employed by the communicating hosts. Hence, the adversary cannot subvert the cryptographic algorithms. We similarly require that in the presence of secure SCADA protocols, our NIDS will be configured with the necessary cryptographic keys so it would be able to decrypt the examined traffic.

\begin{figure}[t]
\centering
  \includegraphics [ width=0.85\textwidth, natwidth=500,natheight=60,trim = 50 20 0 0]
   {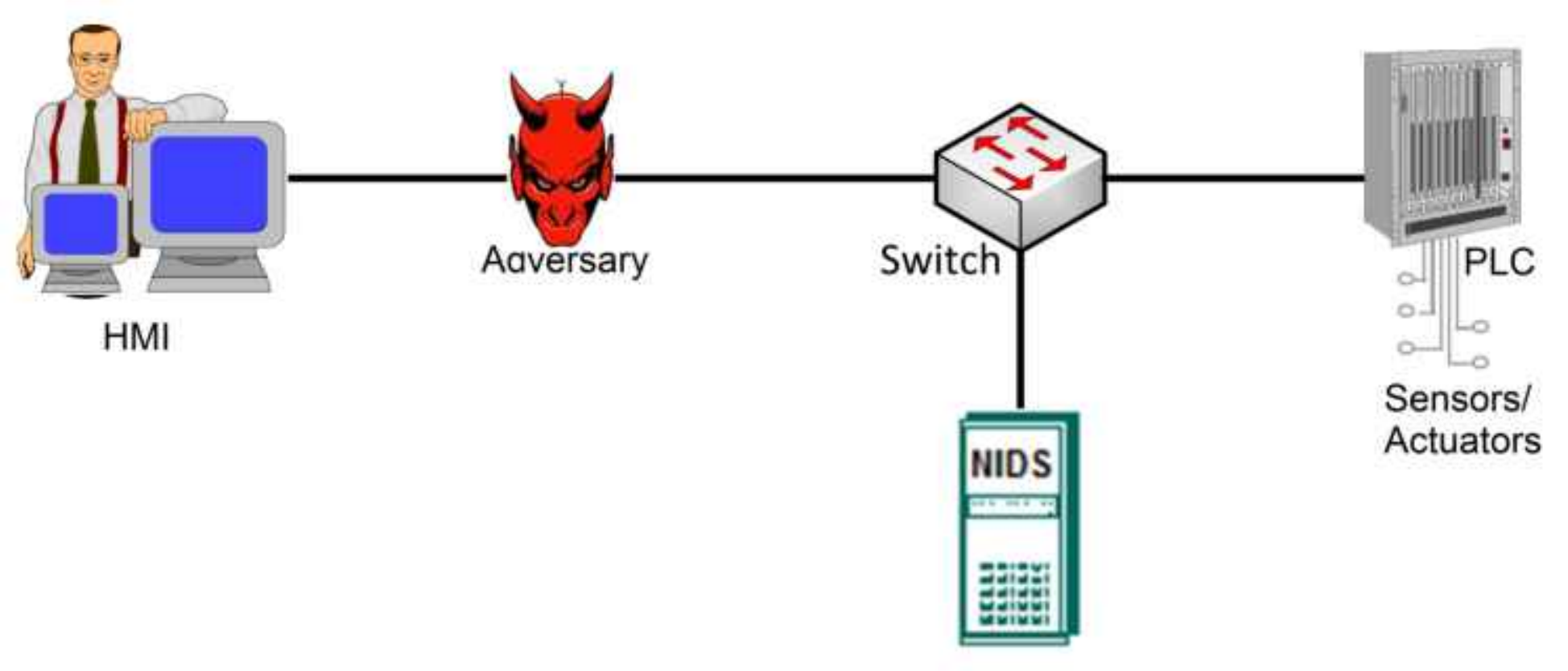}
  \hfill
  \caption{Placing the Network Intrusion Detection System (NIDS) in a SCADA network}
  \label{fig:systemDiagram}
\end{figure}

Our sensor would be located in the network segment where it can passively monitor traffic that was already modified by the adversary and just before the PLC as illustrated in Figure \ref{fig:systemDiagram}. The sensor is not located inline so it does not affect the normal network operation (e.g., port mirroring or similar mechanism is used to instruct the switch to send copies of network traffic to the NIDS).

We further assume that the adversary has in-depth knowledge of the architecture of the SCADA network and the various PLCs as well as sufficient knowledge of the physical process and the means to manipulate it via the SCADA system. Thus the adversary has the ability to fabricate messages that would result in real-world physical damage. Fovino et al.\ illustrated such an attack by describing a system with a pipe in which high pressure steam flows \cite{fovino2010modbus}. The pressure is regulated by two valves. An attacker capable of sending packets to the PLCs can force one valve to complete closure, and force the other to open. Each of these ICS commands is perfectly legal when considered individually, however when sent in a certain order they  bring the system to a critical state. \cite{RobertT} presents an attack scenario where a system-wide water hammer effect is caused simply by opening or closing major control valves too rapidly. This can result in a large number of simultaneous main breaks.

Digital attacks that cause physical destruction of equipment do occur. Perhaps most notably is the attack on an Iranian nuclear facility in 2010 (Stuxnet) to sabotage centrifuges at a uranium enrichment plant. The Stuxnet malware \cite{falliere2011w32,langner2011stuxnet} implemented a water--hammer attack by changing centrifuge operating parameters in a pattern that damaged the equipment -- while sending normal status messages to the HMI to hide the fact that an attack is under way. \cite{SicherheitDeutschland} describes a more recent event, where hackers had struck an unnamed steel mill in Germany, by manipulating and disrupting control systems to such a degree that a blast furnace could not be properly shut down, resulting in “massive”-though unspecified-damage.

Fundamentally all these attacks work by injecting messages into the communication stream---possibly legitimate messages---on an attacker-selected pattern and schedule. Hence a good anomaly detection system needs to model not only the messages in isolation but also their sequence and timing.

Note that our anomaly detection approach does not distinguish between malicious events and faulty events.

\section{A Statechart-based Solution} \label{sec:Statechart}
\begin{figure}[t]
\centering
   \includegraphics [ width=0.75\textwidth, natwidth=500,natheight=60, trim = 30 20 0 0]
    {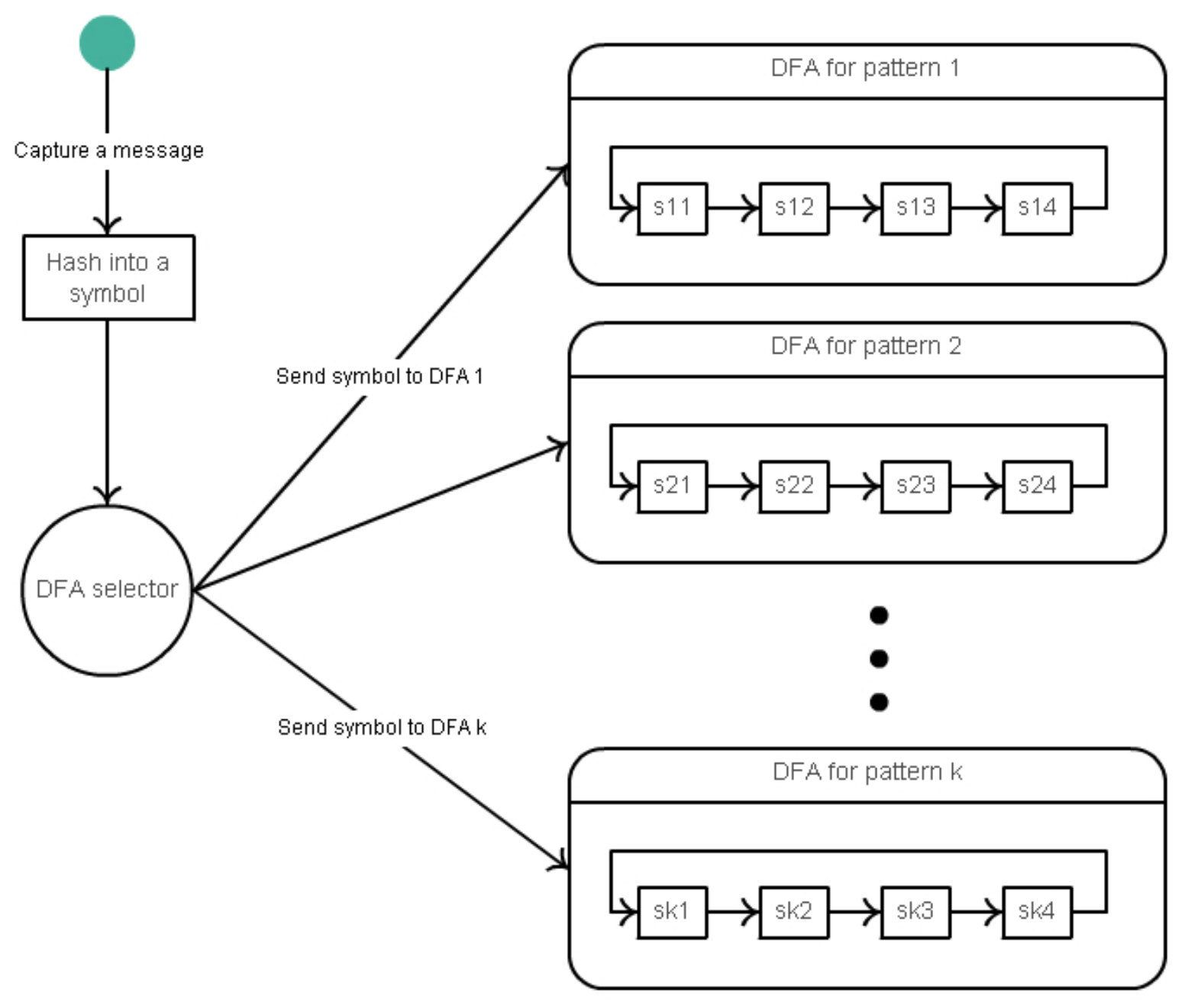}
   \hfill
   \caption{A Statechart DFA model}
   \label{fig:statechart}
\end{figure}

Our first observation is that, as hypothesized by \cite{Caselli} modern HMIs employ thread-based architecture (e.g., this is how the Afcon's Pulse HMI \cite{AfconPulse} is built):
While each thread is responsible for certain tasks (e.g., controlling access to a range of registers on a PLC), the threads run concurrently with different scheduling frequencies, and share the same network connections.
Hence, to accurately model the traffic produced by such an HMI (with the PLC's responses), we should use a formalism that is more descriptive than a basic DFA. Our choice is to base our model on the concept of a Statechart \cite{Harel1987}: the periodic traffic pattern driven by each thread in the HMI is modeled by its own DFA within the {\em Statechart}, see Figure \ref{fig:statechart}.
Each DFA is built using the unsupervised learning stage of the GW model. The {\em Statechart} also contains a DFA-selector to switch between DFAs.

\subsection{The Statechart Enforcement Phase}
During the enforcement stage, each DFA in
the {\em Statechart} maintains its own state, from which it transitions based on the observed symbols (messages).

The DFA-selector's role is to send the input symbol $s$ to the appropriate DFA. To do so it relies on a symbol-to-DFA mapping $\phi$: $\phi(s)$ denotes the set of DFAs that have symbol $s$ in their pattern. If each pattern has a unique set of symbols then $\phi$ is 1-1. However, in the general case, a symbol may appear in multiple patterns and $\phi$ is one-to-many. Upon receiving a symbol $s$ the DFA-selector uses the following algorithm:
\begin{itemize}
\item If $\phi(s) = \varnothing$ the DFA-selector reports an ``Unknown'' symbol.
\item If $\phi(s) = \{D\}$, i.e., the symbol is a unique symbol of a single  DFA $D$, then $s$ is sent to $D$, which handles it using its own transition function.
\item Else, if $|\phi(s)|>1$, the selected DFA is the member of $\phi(s)$ for which the absolute difference between the current time and the {\em predicted arrival time of $s$} is minimal.
\end{itemize}
In order to implement this policy:
\begin{itemize}
\item During the DFA learning stage of the GW model, for each state $r$ in the DFA's pattern we calculate the average time difference to its immediate successor in the cyclic pattern (along the ``Normal'' transition). We denote this Time to Next State by $TNS(r)$.
\item During the enforcement phase, each DFA $D$ retains the time-stamp $T_{last}(D)$ of the last symbol that was processed by it (in addition to the identifier of the current state).
\end{itemize}

The predicted arrival time $T_{pred}(s,D)$ of a symbol $s$ for a DFA $D \in \phi(s)$ which is currently at state $q$, is calculated as follows:
\begin{enumerate}
\item Identify the tentative state $q'$ that DFA $D$ transitions to from state $q$ upon symbol $s$. Note that $q'$ is not necessarily the immediate successor of $q$ in the pattern---the transition from $q$ to $q'$ may be a ``Miss'' or a ``Retransmission''.
\item Let $P(q,q')$ denote the path of DFA states starting at $q$ and ending at $q'$ along the ``Normal'' transitions (not including $q'$). Then  $T_{pred}(s,D) = T_{last}(D) + \sum_{r\in P(q,q')}TNS(r)$: The predicted arrival time is the sum of inter-symbol delays along the ``Normal'' path between $q$ and the tentative transition-to state $q'$ added to the time-stamp of the last symbol processed by DFA $D$.
\end{enumerate}

\subsection{The Statechart Unsupervised Learning Phase}

The goal of the learning phase is to construct the {\em Statechart} for a specific HMI-PLC channel, given a captured stream of symbols from the channel. For this we need to create the  symbol-to-DFA mapping $\phi$, for the use of the DFA selector, and we need to create the individual DFAs themselves.
In \cite{KleinmannWool2015} we treated the simple cases and constructed the {\em Statechart} as follows:
\begin{enumerate}
\item Split the channel's input stream into multiple sub-channels.
\item For each sub-channel use the GW unsupervised learning algorithm to create a DFA.
\item Create the DFA-selector's mapping $\phi$ from the
sub-channel DFAs.
\end{enumerate}

\begin{figure}[t]
\centering
   \includegraphics [ width=0.65\textwidth, natwidth=500,natheight=60, trim = -30 20 0 0]
    {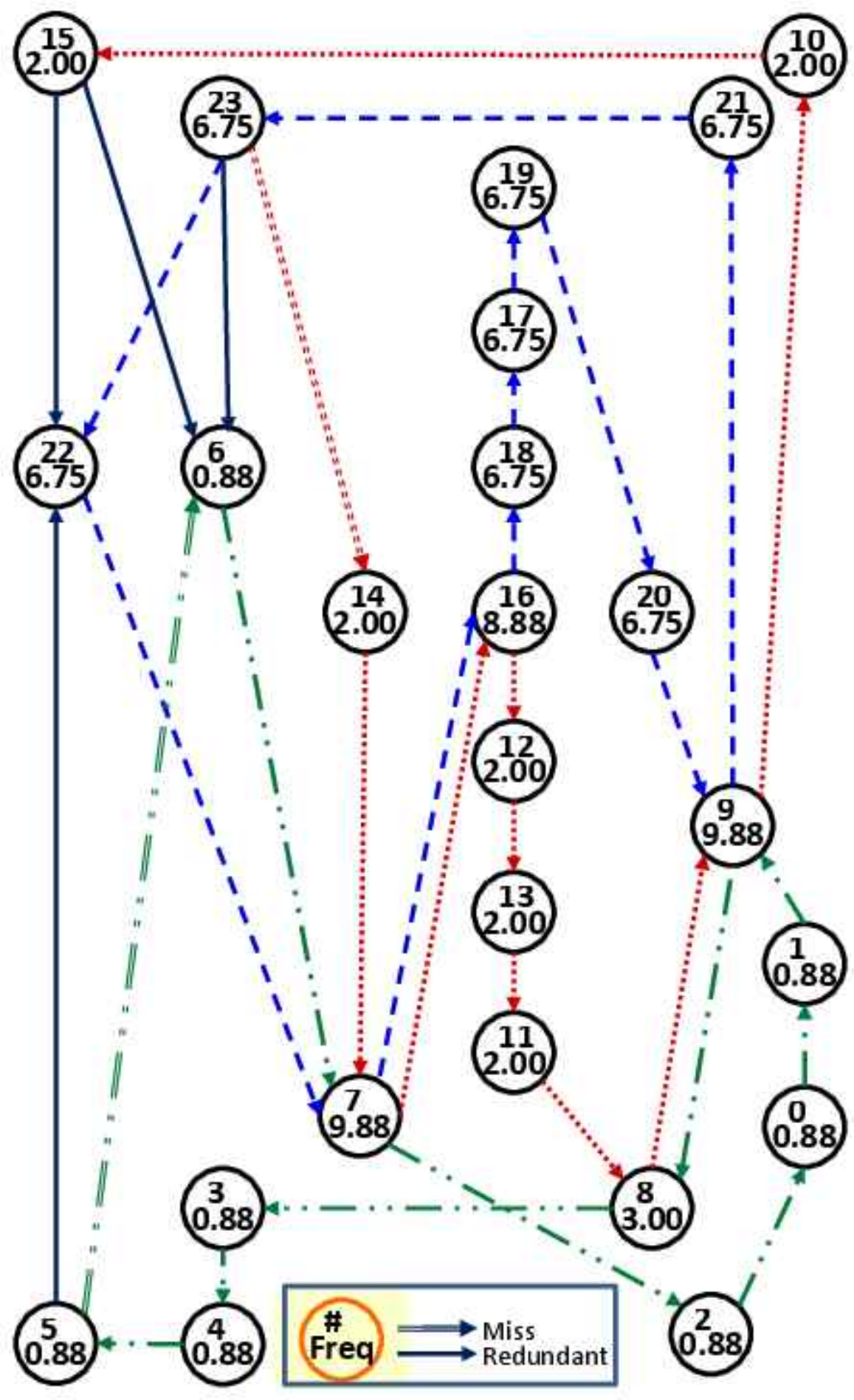}
   \hfill
   \caption{A DTMC of SCADA Traffic (47,473 symbols captured along 1000 sec). The node frequencies are in thousands, for the first 16.82 sec captured trace comprised of 800 symbols.}
   \label{fig:dtmc}
\end{figure}

The sub-channel splitting (step 1) can be implemented when we know how many sub-channels can exist, each sub-channel has a unique set of symbols, and there is a filter criterion to recognize them. However, a difficult case is when we don't know in advance how many sub-channels exist, and the sub-channels potentially have overlapping symbols.
In this paper we introduce a new sub-channel splitting algorithm that performs well even in the more challenging cases. It uses a Deterministic Time Markov Chain (DTMC) and graph theory concepts to create a DFA, or several DFA alternatives, for each cyclic pattern in a PLC-HMI channel. Then, for each PLC-HMI channel, a statechart is built out of any combination of the multiple DFAs and the best-performing combination is used.

\subsection{Building a DTMC}
For our purposes a Discrete Time Markov Chain (DTMC) $M$ is a tuple $(S,P)$ where:
\begin{itemize}
\item S is a countable nonempty set of states
\item P: S x S $\to [0,1]$, is the transition probability function s.t., $\Sigma_{s'} P(s,s') = 1$ :
P(s,s') is the probability to jump from $s$ to $s'$ in one step
\end{itemize}

A state graph of DTMC $M$ is a directed graph G = (V,E) where vertices in V are states of $M$, and $(s, s') \in$ E iff  $P(s,s') > 0$. Henceforth we refer to the state graph of the DTMC simply by DTMC.

In our model the nodes V represent the symbols (messages) observed in the SCADA trace. A directed edge e = (u,v) represents a ``followed-by'' relation: symbol u is followed by symbol v in the learned trace. For e=(u,v) the probability P(e) represents the fraction of time that the transition u$\to$v was traversed, out of all transitions exiting u.

We also assign a frequency freq(v) to the nodes v$\in$V, representing the total number of time that v was observed in the learned trace.

Following \cite{Goldenberg201363,KleinmannWjdfsl}, in order to construct the model's DFAs and Statechart, there is a need to define the symbols.
\cite{Goldenberg201363} defined a symbol for Modbus as a concatenation of the message type, function code, and address range, totaling 33-bits; \cite{KleinmannWjdfsl} selected several PDU fields of the S7-0x32 protocol, that were then hashed into 64-bit symbols.

After selecting which fields of a SCADA message constitute a symbol in the DFA's alphabet, we generate a symbol out of each SCADA message that is captured.
We count the number of occurrences of each symbol as well as the number of occurrences of each bigram (contiguous sequence of 2 symbols).

We then construct the DTMC using the symbols as the DTMC states, and defining the transition probability by normalizing the bigram count over all edges exiting a symbol. The construction of a DTMC is independent of the SCADA protocol and is completely automated. Figure \ref{fig:dtmc} shows a DTMC of SCADA traffic (that was constructed according to scenario \#6 of the sequences used to generate the synthetic datasets, see Table \ref{tab:sequence-attributes-for-synthetic-data}). Each circle in the Figure represents a vertex of the DTMC graph with its id number and its occurrence frequency. A DTMC edge is represented by an arrow between two circles.

We noticed that in our data there are bigrams that occur rarely and represent transitions between different cycles rather than transition between symbols of a certain cycle. This happens due to the multiplexing of the different cycles within the traffic stream. Consequently, we decided to filter out bigrams that occur rarely, i.e., bigrams that occur below a certain threshold of the maximum occurrences for the target vertex. We used a threshold $T_{rare}$ of 10\%.

\begin{table}%
\caption{Overview of the sets of sequences used to generate the synthetic datasets\label{tab:sequence-attributes-for-synthetic-data}}{%
\resizebox{\columnwidth}{!}{%
\begin{tabular}{|c|c|c|c|cccccccccc}
\hline
\cline{1-4} \cline{6-9} \cline{11-14}
\textbf{ID}                 & \textbf{Len.} & \textbf{Uniq.} & \textbf{Period} & \multicolumn{1}{l|}{\textbf{}} & \multicolumn{1}{l|}{\textbf{ID}}                  & \multicolumn{1}{l|}{\textbf{Len.}} & \multicolumn{1}{l|}{\textbf{Uniq.}} & \multicolumn{1}{l|}{\textbf{Period}} & \multicolumn{1}{l|}{\textbf{}} & \multicolumn{1}{l|}{\textbf{ID}}                  & \multicolumn{1}{l|}{\textbf{Len.}} & \multicolumn{1}{l|}{\textbf{Uniq.}} & \multicolumn{1}{l|}{\textbf{Period}} \\ \cline{1-4} \cline{6-9} \cline{11-14}
\multirow{2}{*}{\textbf{1}} & 6               & 6              & 300             & \multicolumn{1}{c|}{}          & \multicolumn{1}{c|}{\multirow{3}{*}{\textbf{7}}}  & \multicolumn{1}{c|}{10}              & \multicolumn{1}{c|}{8}              & \multicolumn{1}{c|}{300}             & \multicolumn{1}{c|}{}          & \multicolumn{1}{c|}{\multirow{4}{*}{\textbf{11}}} & \multicolumn{1}{c|}{10}              & \multicolumn{1}{c|}{8}              & \multicolumn{1}{c|}{250}             \\ \cline{2-4} \cline{7-9} \cline{12-14}
                            & 4               & 4              & 950             & \multicolumn{1}{c|}{}          & \multicolumn{1}{c|}{}                             & \multicolumn{1}{c|}{8}               & \multicolumn{1}{c|}{7}              & \multicolumn{1}{c|}{350}             & \multicolumn{1}{c|}{}          & \multicolumn{1}{c|}{}                             & \multicolumn{1}{c|}{4}               & \multicolumn{1}{c|}{2}              & \multicolumn{1}{c|}{650}             \\ \cline{1-4} \cline{7-9} \cline{12-14}
\multirow{2}{*}{\textbf{2}} & 6               & 6              & 300             & \multicolumn{1}{c|}{}          & \multicolumn{1}{c|}{}                             & \multicolumn{1}{c|}{10}              & \multicolumn{1}{c|}{9}              & \multicolumn{1}{c|}{400}             & \multicolumn{1}{c|}{}          & \multicolumn{1}{c|}{}                             & \multicolumn{1}{c|}{6}               & \multicolumn{1}{c|}{4}              & \multicolumn{1}{c|}{1100}            \\ \cline{2-4} \cline{6-9} \cline{12-14}
                            & 4               & 4              & 950             & \multicolumn{1}{c|}{}          & \multicolumn{1}{c|}{\multirow{3}{*}{\textbf{8}}}  & \multicolumn{1}{c|}{10}              & \multicolumn{1}{c|}{8}              & \multicolumn{1}{c|}{300}             & \multicolumn{1}{c|}{}          & \multicolumn{1}{c|}{}                             & \multicolumn{1}{c|}{8}               & \multicolumn{1}{c|}{7}              & \multicolumn{1}{c|}{420}             \\ \cline{1-4} \cline{7-9} \cline{11-14}
\multirow{2}{*}{\textbf{3}} & 6               & 4              & 300             & \multicolumn{1}{c|}{}          & \multicolumn{1}{c|}{}                             & \multicolumn{1}{c|}{8}               & \multicolumn{1}{c|}{7}              & \multicolumn{1}{c|}{850}             & \multicolumn{1}{c|}{}          & \multicolumn{1}{c|}{\multirow{4}{*}{\textbf{12}}} & \multicolumn{1}{c|}{6}               & \multicolumn{1}{c|}{4}              & \multicolumn{1}{c|}{250}             \\ \cline{2-4} \cline{7-9} \cline{12-14}
                            & 4               & 1              & 400             & \multicolumn{1}{c|}{}          & \multicolumn{1}{c|}{}                             & \multicolumn{1}{c|}{10}              & \multicolumn{1}{c|}{9}              & \multicolumn{1}{c|}{1300}            & \multicolumn{1}{c|}{}          & \multicolumn{1}{c|}{}                             & \multicolumn{1}{c|}{4}               & \multicolumn{1}{c|}{4}              & \multicolumn{1}{c|}{350}             \\ \cline{1-4} \cline{6-9} \cline{12-14}
\multirow{2}{*}{\textbf{4}} & 6               & 4              & 300             & \multicolumn{1}{c|}{}          & \multicolumn{1}{c|}{\multirow{3}{*}{\textbf{9}}}  & \multicolumn{1}{c|}{10}              & \multicolumn{1}{c|}{7}              & \multicolumn{1}{c|}{300}             & \multicolumn{1}{c|}{}          & \multicolumn{1}{c|}{}                             & \multicolumn{1}{c|}{10}              & \multicolumn{1}{c|}{9}              & \multicolumn{1}{c|}{550}             \\ \cline{2-4} \cline{7-9} \cline{12-14}
                            & 4               & 2              & 950             & \multicolumn{1}{c|}{}          & \multicolumn{1}{c|}{}                             & \multicolumn{1}{c|}{8}               & \multicolumn{1}{c|}{4}              & \multicolumn{1}{c|}{350}             & \multicolumn{1}{c|}{}          & \multicolumn{1}{c|}{}                             & \multicolumn{1}{c|}{8}               & \multicolumn{1}{c|}{7}              & \multicolumn{1}{c|}{420}             \\ \cline{1-4} \cline{7-9} \cline{11-14}
\multirow{3}{*}{\textbf{5}} & 10              & 9              & 300             & \multicolumn{1}{c|}{}          & \multicolumn{1}{c|}{}                             & \multicolumn{1}{c|}{10}              & \multicolumn{1}{c|}{8}              & \multicolumn{1}{c|}{400}             & \multicolumn{1}{c|}{}          & \multicolumn{1}{c|}{\multirow{4}{*}{\textbf{13}}} & \multicolumn{1}{c|}{10}              & \multicolumn{1}{c|}{9}              & \multicolumn{1}{c|}{300}             \\ \cline{2-4} \cline{6-9} \cline{12-14}
                            & 4               & 2              & 600             & \multicolumn{1}{c|}{}          & \multicolumn{1}{c|}{\multirow{3}{*}{\textbf{10}}} & \multicolumn{1}{c|}{6}               & \multicolumn{1}{c|}{3}              & \multicolumn{1}{c|}{300}             & \multicolumn{1}{c|}{}          & \multicolumn{1}{c|}{}                             & \multicolumn{1}{c|}{4}               & \multicolumn{1}{c|}{2}              & \multicolumn{1}{c|}{600}             \\ \cline{2-4} \cline{7-9} \cline{12-14}
                            & 4               & 3              & 200             & \multicolumn{1}{c|}{}          & \multicolumn{1}{c|}{}                             & \multicolumn{1}{c|}{4}               & \multicolumn{1}{c|}{2}              & \multicolumn{1}{c|}{350}             & \multicolumn{1}{c|}{}          & \multicolumn{1}{c|}{}                             & \multicolumn{1}{c|}{4}               & \multicolumn{1}{c|}{2}              & \multicolumn{1}{c|}{200}             \\ \cline{1-4} \cline{7-9} \cline{12-14}
\multirow{3}{*}{\textbf{6}} & 10              & 7              & 300             & \multicolumn{1}{c|}{}          & \multicolumn{1}{c|}{}                             & \multicolumn{1}{c|}{6}               & \multicolumn{1}{c|}{2}              & \multicolumn{1}{c|}{400}             & \multicolumn{1}{c|}{}          & \multicolumn{1}{c|}{}                             & \multicolumn{1}{c|}{6}               & \multicolumn{1}{c|}{3}              & \multicolumn{1}{c|}{350}             \\ \cline{2-4} \cline{6-9} \cline{11-14}
                            & 10              & 7              & 950             &                                &                                                   &                                      &                                     &                                      &                                &                                                   &                                      &                                     &                                      \\ \cline{2-4}
                            & 10              & 7              & 2000            &                                &                                                   &                                      &                                     &                                      &                                &                                                   &                                      &                                     &                                      \\ \cline{1-4}

\end{tabular}}
}
\end{table}%

\subsection{Identifying a Symbol Set for each Cycle}
In order to identify the set of symbols for each cyclic pattern out of the DTMC, we designed and applied an iterative algorithm. The major steps of the algorithm are depicted in Algorithm \ref{Algo}. The algorithm starts by selecting all the vertices with only one incoming edge and one outgoing edge. The motivation to start with these vertices is based on the assumption that each of these vertices belongs only to one set. Attached to each vertex is the occurrence frequency of its associated SCADA symbol $v_{freq}$. Consequently, it is assumed that $v_{freq}$ is the frequency of the cycle to which that symbol belongs to. These vertices are classified to sets where each set contains vertices of similar frequency F. A vertex is added to a certain set S if its frequency is ``similar'' (denoted by $\simeq$) to the frequency of the set. In the algorithm we use the $\simeq$ operator to denote frequency extended with a threshold $T_{sim}$, i.e., $S_{frequency}\times(1-T_{sim}) < v_{freq} < S_{frequency}\times(1+T_{sim})$, where $T_{sim}$ is a configurable threshold. In our experiments we used a threshold of $T_{sim}$=0.05. The first vertex $V$ that does not belong to any of the existing sets ``creates'' a new set with a frequency $v_{freq}$.

A vertex can be a member of multiple sets, signifying the occurrence of a symbol in multiple cyclic patterns. We call each of these memberships - another instance of the vertex.

In steps 2 - 5 vertices that do not yet belong to any set (or that have some instances that have not been classified yet) are added to existing sets. In steps 2 vertices that have only one incoming edge or outgoing edge as well as a frequency that is similar to a frequency of one of the existing sets, are added to the corresponding set. In steps 3 vertices whose frequency is the sum of their adjacent (connected by an entering edge or by an exiting edge) vertices (as long as each of the adjacent vertices belongs to only one set) are added to the corresponding sets of the adjacent vertices. In step 4 several combinations are created out of adjacent vertices whose frequencies sum up to the vertex frequency. When more than one combination is created, each one of them is considered separately and multiple symbol-set alternatives may be created for the specific pattern cycle. In step 5 each of the remaining vertices is examined. If a vertex has both an incoming edge and an outgoing edge from/to the same set, and one of the adjacent vertices that belongs to that set has both in-degree = 1 and out-degree = 1, then the vertex is added to that set.

Note that in steps 2 - 5 new sets are not created. Instead the remaining vertices are classified into the existing sets. An exception is the 6th step where a new set is formed for vertices that could not be classified to any other set. The frequency of this set is fixed to the minimal frequency among its vertex members. Then, in the last step (step \#7) this frequency is deducted from each of the vertices that are members of the new set, and the algorithm tries to classify additional instances of those vertices with remaining frequency $F > 0$.

\begingroup
\renewcommand{\arraystretch}{0.85}
\begin{pseudocode}{Get Symbol Sets for Cycles}{\ }
\label{Algo}

Let\ v\ represent\ a\ vertex,\ S\ represent\ set\ of\ vertices\\
\textbf{\underline{Step 1: Basic in and out degree 1}}\\
\forall i,v_i \mid v_i.inDegree = 1, v_i.outDegree = 1,\\
\IF \exists S_j \mid S_j.freq \simeq v_i.freq
\THEN
S_j \GETS S_j \cup \{v_i\}
\ELSE
k \GETS max(j)+1,\ S_k \GETS \{v_i\}\\
\textbf{\underline{Step 2: Additional degree 1}}\\
RemainVrtx \GETS \{\forall i, \forall j, v_i \mid v_i\not\in S_j \}\\
\FOREACH v \in RemainVrtx \DO
\BEGIN
\IF (v.inDegree = 1 \OR v.outDegree = 1)\\
\-\hspace{0.3cm} \AND \exists S_j \mid S_j.freq \simeq v.freq
\THEN
S_j \GETS S_j \cup \{v\}
\END\\
\textbf{\underline{Step 3: Unique adjacents }}\\
RemainVrtx \GETS \{\forall i, \forall j, v_i \mid v_i\not\in S_j \}\\
\FOREACH v \in RemainVrtx \DO
\BEGIN
V_{in} \GETS \{\forall i, v_i \mid (v_i,v)\ is\ an\ incoming\ edge\\
\-\hspace{2.1cm} \ of\ v, v_i \in S_j, v_i \not\in S_k, k\ne j\}\\
\IF  (\sum_{\forall S_l \in V_{in}} S_l.freq) \simeq v.freq
\THEN
\FOREACH S_l \in V_{in} \ \ S_l \GETS S_l \cup \{v\}
\ELSE
\BEGIN
V_{out} \GETS \{\forall i, v_i \mid (v,v_i)\ is\ an\ outgoing\\
\-\hspace{1cm} edge\ of\ v, v_i \in S_j, v_i \not\in S_k, k\ne j\}\\
\IF  (\sum_{\forall S_l \in V_{out}} S_l.freq) \simeq v.freq
\THEN\\
\-\hspace{0.7cm}\FOREACH S_l \in V_{out} \ \ S_l \GETS S_l \cup \{v\}
\END
\END\\
\textbf{\underline{Step 4: Permutations of adjacents}}\\
RemainVrtx \GETS \{\forall i, \forall j, v_i \mid v_i\not\in S_j \}\\
V_{adj} \GETS \{\forall i, v_i \mid ((v_i,v)\ is\ an\ incoming\ edge\ of\ v) \OR \\
\-\hspace{2.3cm}((v,v_i)\ is\ an\ outgoing\ edge\ of\ v)\}\\
Find\ all\ V_{{adj}_p} \mid V_{{adj}_p} \subseteq V_{adj},\\
\-\hspace{1.5cm} (\sum_{\forall S_l \in V_{{adj}_p}} S_l.freq) \simeq v.freq\\
\FOREACH V_{{adj}_p}
\BEGIN
Keep\ a\ permutation\ of\ all\ its\ sets\\
\FOREACH S_l \in V_{{adj}_p} \ \ S_l \GETS S_l \cup \{v\}
\END\\
\textbf{\underline{Step 5: Include ''via-vertices''}}\\
RemainVrtx \GETS \{\forall i, \forall j, v_i \mid v_i\not\in S_j \}\\
\FOREACH S_j
\BEGIN
\IF v_i \in S_j \AND v_j \in S_j \AND ((v_i,v)\ is\\
 an\ ingoing\ edge\ of\ v) \AND (v,v_j)\ is\\
 an\ outgoing\ edge\ of\ v \AND\\
 ((v_i.inDegree = 1 \AND v_i.outDegree = 1) \OR\\
(v_j.inDegree = 1 \AND v_j.outDegree = 1))
\THEN
\BEGIN
v.freq \GETS v.freq - S_j.freq\\
S_j \GETS S_j \cup \{v\}
\END
\END\\
\textbf{\underline{Step 6: Collect all remains in a separate set}}\\
RemainVrtx \GETS \{\forall i, \forall j, v_i \mid v_i\not\in S_j \}\\
k \GETS max(j)+1,\ S_k \GETS RemainVrtx\\
MinFreq \GETS Min_{\forall i, v_i \in S_k}(v_i.freq)\\
\FOREACH v_i \in S_k  \\
\-\hspace{0.8cm}S_i.freq \GETS v_i.freq - MinFreq\\
\textbf{\underline{Step 7: Add remained instances to existing sets}}\\
RemainVrtx \GETS \{\forall i, v_i \mid v_i \in RemainVrtx, v_i.freq \not\simeq 0 \}\\
\FOREACH v \in RemainVrtx \DO
\BEGIN
\IF
\exists S_j \mid S_j.freq \simeq v.freq
\THEN
S_j \GETS S_j \cup \{v\}
\END\\
\end{pseudocode}

\endgroup

\subsection{Constructing the Cycles}
Once we identified the symbol sets for each cyclic pattern, we construct a subgraph per each cyclic pattern. We add to each subgraph every identified symbol as a vertex and every DTMC edge between symbols of the cycle as an edge.
After a subgraph is constructed, we check whether it contains an Euler cycle/s by testing the requirement that each of its vertices has an in-degree that is equal to its out-degree.
In case there are vertices with an in-degree that is different from the out-degree, we try to ``fix'' the graph by dropping redundant edges and/or adding a missing edge as follows:

\begin{itemize}
\item We use the observation that the SCADA messages of each cycle are sent in a burst, followed by a long delay (during which other cycles can appear). Therefore sometime the edge between the last message of the cycle and the first message of the cycle will be absent from the DTMC and consequently missing in the constructed subgraph of the pattern cycle. We call such an edge a {\em missed edge}.
\item In some cases the resulting DTMC would includes edges which we call {\em redundant edges}. An edge is considered a {\em redundant edge} if it connects vertices that represent symbols that are, in turn, representing SCADA messages that are \textbf{not} sent sequentially within a certain cyclic pattern. In addition the DTMC may include edges between vertices that are not unique to specific cyclic pattern, i.e., vertices that occur in several cyclic patterns. These edges may represent SCADA messages that are sent sequentially within a certain cyclic pattern but may \textbf{not} be sent sequentially within other cyclic pattern/s even though these cyclic patterns include the same edge vertices. Consequently these edges are redundant edges in those other patterns.
\end{itemize}

Therefore, in order to construct the different cycles, we examine all the vertices with an in-degree that is not equal to their out-degree, and we take the following actions:
\begin{itemize}
\item In order to add missed edges in a certain subgraph, we look for a pair of vertices (of the subgraph) where one has an in-degree greater by one than its out-degree, while the other has an out-degree greater by one than its in-degree. We add an edge from the vertex with a missing outgoing edge to the vertex which lacks one incoming edge.
\item In order to eliminate redundant edges for a certain subgraph, we calculate the median of the number of edge (bigram) occurrences for that subgraph. Then we drop all the edges of the subgraph with number of occurrences that is below (or above) the median edge occurrences (minus, or plus, respectively, of a certain threshold. In our experiments we used a threshold of 5\%).
\end{itemize}

The addition of missed edges and the removal of redundant edges are illustrated in Figure \ref{fig:dtmc}. This Figure describes the construction of the three cycles of scenario \#6 (see Table \ref{tab:sequence-attributes-for-synthetic-data}). It depicts three subgraphs. The cycles of each of the subgraphs are denoted by the type and color of its arrow lines: dashed blue, dotted red, and dotted-dashed green. The redundant edges are denoted by arrow-lines that are solid and black.
The missed edges are denoted by double lines.

\begin{figure}[t]
\centering
   \includegraphics [ width=0.65\textwidth, natwidth=500,natheight=60, trim = 0 10 0 0]
    {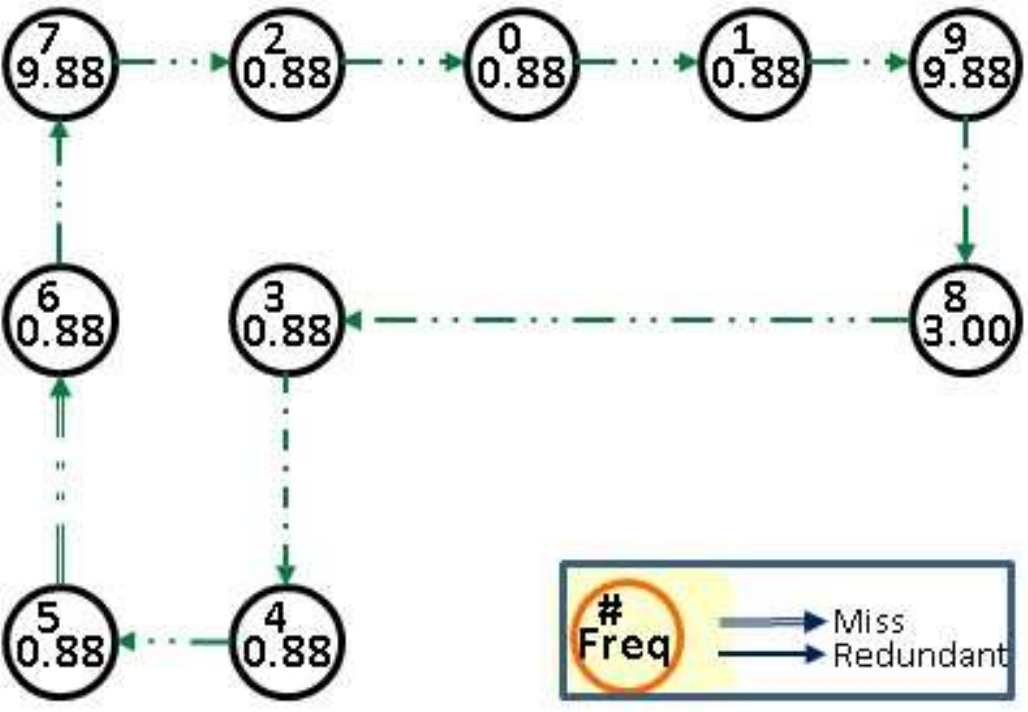}
   \hfill
\caption{One of the detected cycles of scenario \#6, for a set with frequency 0.88}
    \label{fig:dtmc1}
\end{figure}

Figure \ref{fig:dtmc1} is derived from Figure \ref{fig:dtmc}. It shows one of the three cycles, that consists of 10 vertices and 9 edges. Our algorithm successfully detects the missing edge ($5 \to 6$) and adds it to the graph in order to enable the discovery of an Eulerian cycle.

In order to find Euler cycle/s we implemented Hierholzer's algorithm \cite{hierholzer1873moglichkeit}. In certain cases several Euler cycles can be found for a given subgraph. Consequently, we keep all the discovered cycles and create several combinations of the discovered cycles of different sub-graphs. For each combination we create a corresponding Statechart. We then run the DFA learning validation process for each of the Statecharts separately, and decide which Statechart to select for the enforcement stage, based on the best validation result.

\subsection{Deducing the Time Gaps}
The last thing we need for the Statechart is information on the average time gap between successive symbols in each of the cyclic patterns, i.e., for each state $r$ in the DFA's pattern we need to calculate the average time difference to its immediate successor in the cyclic pattern (along the ``Normal'' transition). This is needed since during the enforcement phase, each DFA $D$ retains the time-stamp $T_{last}(D)$ of the last symbol that was processed by it (in addition to the identifier of the current state). In cases of symbol overlap between different patterns, it is this time gap information that helps the Statechart to decide on the next state to move to. The time gap between successive symbols of the same pattern is usually very small except the time gap between the last symbol of the pattern and the first symbol of the pattern (of the next burst). The time difference between an occurrence of the first symbol after the last symbol, and its next occurrence after the last symbol, equals to the time period of the pattern cycle.

\begin{figure}[t]
\centering
   \includegraphics [ width=0.95\textwidth, natwidth=500,natheight=60, trim = 30 300 0 250]
    {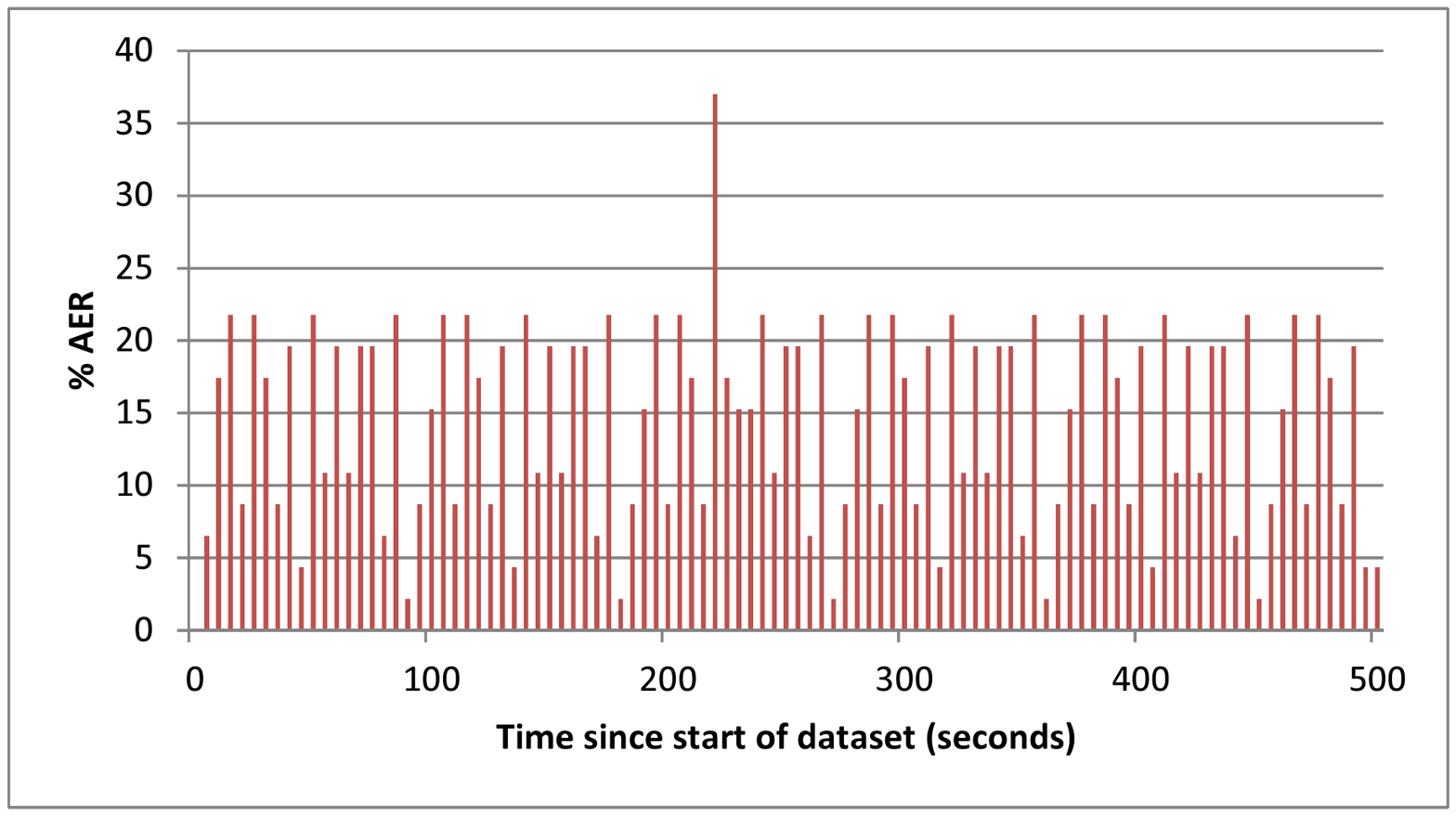}
   \hfill
\caption{One of the detected cycles of scenario \#6, for a set with frequency 0.88}
    \label{figure:S7-0x72-abnormals-dataset1}
\end{figure}

\section{The S7-0x72 Protocol}
{\bf The S7 PLC Platform.}
The Siemens SIMATIC S7 product line is estimated to have over 30\% of the worldwide PLC market \cite{plcmarket}. It includes both standard PLC models (S7-200, S7-300 and S7-400), and new generation PLCs (S7-1200 and S7-1500).
Siemens has its own HMI software for its SIMATIC products called STEP7 and uses its own S7 communication protocol, over TCP port 102.

Two different protocol flavours are implemented by SIMATIC S7 products: The standard SIMATIC S7 PLCs implement a legacy S7 flavor, identified by the value 0x32, while the new generation PLCs implement a very different S7 flavor identified by 0x72. Among other changes, the newer S7-0x72 protocol also supports security features.

The standard S7-0x32 protocol is quite well understood, and a standard Wireshark dissector is available for it. The newer S7-0x72 protocol is not yet fully described in open literature. There is, however, a Wireshark dissector for it which is still in beta status \cite{s7dissector}.

A unique feature of the S7-0x72 protocol is its optional subscription model (in addition to the traditional request-response pattern). The HMI can send a special ``subscribe'' message, referring to certain control variables, to a PLC. Subsequently the PLC sends back a periodic stream of ``notification'' messages with the values of the subscribed variables. The challenge that this subscription model poses to a DFA-based anomaly detection system is that the notification messages are sent asynchronously, and are not part of the HMI-driven  request-response pattern.

\noindent{\bf Experimenting with the S7-0x72 Data.}
Due to the proprietary nature and potential sensitivity of SCADA operations, real SCADA network data is rarely released to researchers. An important aspect of this work is that we were able to collect and analyze traces from a production S7 network running the S7-0x72 protocol from a control network of a solar power plant.

In these traces we observed a single channel between the HMI and a Siemens S7-1500 PLC. We observed both the request-response and the unique subscribe/notification communication patterns. An overview of the S7 datasets can be found in Table \ref{table:s7result}. During our recordings the infrastructure was running normally without any intervention of operators. 

The message format and protocol semantics described here are based on the reverse engineering work of Wiens \cite{s7dissector}. Somewhat surprisingly the S7-0x72 message formats are very different from those of the older S7-0x32 protocol, even though the overall protocol semantics are quite similar. An S7 0x72 packet is composed of the following parts:
\begin{itemize}
\item Header: `magic ID' byte with a value of 0x72, a PDU type (one byte) and the length of the data part.
\item Data part: includes meta data fields describing the data, data values, and an optional integrity part that is supported only by the newest S7-1500 PLCs (it contains two bytes representing an ID, one byte for the digest length and a 32 byte message digest, which is apparently a cryptographic hash or MAC, details are yet unknown).
\item Trailer: utilized to enable fragmentation.
\end{itemize}

Unlike the packet structure of the S7-0x32 protocol, nearly every field inside the S7-0x72 data part may be composed of recursively defined data structures. Further, elementary components such as numeric values are encoded using the variable-length quantity (VLQ) encoding \cite{vlq}, a universal code that uses an arbitrary number of binary octets.
The precise S7-0x72 packet structure depends on the type of the command and the information it is instructed to carry. The beta Wireshark dissector \cite{s7dissector} is able to parse the structure of over 30 different S7-0x72 commands.

To use the GW model we need to hash the meta-data fields of a SCADA packet into a symbol while ignoring the actual data values. In order to model the S7-0x72 packets we relied on the deep understanding embedded in the Wireshark dissector \cite{s7dissector} to identify the structural meta-data components in the packets (command codes and arguments, register types and reference ids, etc.). In total we extracted 11--17 meta-data fields, comprising of 17--26 bytes, out of typical S7-0x72 packets, which were hashed into 64-bit symbols.

Figure \ref{figure:S7-0x72-abnormals-dataset1} shows the false alarm rate over time of the naive DFA model applied to S7 dataset \#1.
Table \ref{table:s7result} summarizes the results on the two S7 traces, comparing the Naive and Statechart models. We can see that the naive DFA model has high false-alarm rates: 14.54\% and 12.98\%.
The {\em Statechart} model successfully reduced the false-alarm rate by two orders of magnitude, down to at most 0.11\%. The Table shows that the model sizes dropped from the incorrect sizes of 62 and 12 by the naive DFA model down to the correct size of 3 (a request-response pattern of 2 symbols and a notification pattern of 1).

\begin{table}[t]
\hspace{6.75em}
   \begin{tabular}[trim = -100 300 40 245]{|l|c|c|c|c|}
      \hline
      \textbf{Dataset \#}        & \multicolumn{2}{c|}{\textbf{1}}      & \multicolumn{2}{c|}{\textbf{2}}      \\ \hline
      \textbf{Duration}                  & \multicolumn{2}{c|}{560 Sec.}            & \multicolumn{2}{c|}{2632 Sec.}            \\
      \textbf{TCP Packets}                  & \multicolumn{2}{c|}{15875}            & \multicolumn{2}{c|}{67585}            \\
      \textbf{S7 Packets}                  & \multicolumn{2}{c|}{4600}            & \multicolumn{2}{c|}{23553}            \\
      \textbf{AER}                  & \multicolumn{2}{c|}{9.19}            & \multicolumn{2}{c|}{9.16}            \\
      \hline\hline
      \textbf{Dataset \#}        & \multicolumn{2}{c|}{\textbf{1}}      & \multicolumn{2}{c|}{\textbf{2}}      \\ \hline
      \textbf{DFA type}             & \textbf{N} & \textbf{S} & \textbf{N} & \textbf{S} \\ \hline
      \textbf{Model size} & 62             & 3                   & 12             & 3                   \\ \hline
      \textbf{False alrm \%}         & 14.54          & 0.11                & 12.98          & 0                   \\ \hline
      \end{tabular}
      \caption{Results of applying both models}
      \label{table:s7result}
\end{table}

\section{Stress Testing with Synthetic Data}
In the S7-0x72 traces we observed the easy case of sub-channel splitting: the channel consisted of 2 sub-channels, one for request and response messages, and the other for notification messages. Since the message types are in the packet meta-data it is easy to split the input stream. Similarly, \cite{Goldenberg201363} reported that in their Modbus traces the slow and fast cycles had distinct symbols.

However, it seems that in the Modbus data set analyzed by \cite{Caselli} the number of sub-channels is not clear in advance, and sub-channel symbols may be overlapping. Since this data set was not
available to us we chose to stress-test the capabilities of our {\em Statechart} approach in this scenario using synthetic data (see Section \ref{SyntheticData}).

\subsection{Generation of synthetic data}\label{Generation of synthetic data}
 In order to test our model in different scenarios, we implemented a multi-threaded generator, where each of the threads simulates an HMI thread transmitting a cyclic pattern of SCADA commands.
 Each simulated thread has a pattern $P$ of symbols, and a frequency $f$. Every $f$ msec the thread wakes up and  emits the pattern $P$ as a burst, at a 1-msec-per-symbol rate, and returns to sleep. The thread's true timing has a jitter caused by the OS scheduling decisions. Further, when multiple threads are active concurrently then their emitted symbols are arbitrarily serialized.

 The 13 generated scenarios, vary the number of patterns, the number of unique symbols per pattern, and their frequency. the simpler scenarios (1-4) have 2 patterns each - and the most complex multiplex 4 patterns. Table \ref{tab:sequence-attributes-for-synthetic-data} shows the parameters of the scenarios that were used in our simulations.

 In 12 of the 13 scenarios the algorithms were able to split the symbols into sets with 100\% accuracy. Only in one case the algorithms did not identify one of the occurrences of a specific symbol. So in total the algorithms were able to split the symbols into sets with 99.6\% accuracy.

 For the purpose of our evaluation and analysis we defined the following metrics:
 \begin{itemize}
 \item The {\em Symbol Uniqueness} of a channel $= \sum_{i=1}^n U_i / \sum_{i=1}^n L_i$, where
 $L_i$ is the length of the cyclic pattern of sub-channel $i$ and $U_i$ is the number of symbols unique to that sub-channel.
 \item A channel's {\em Time Overlap} is the percentage of 1-msec time slots at which multiple packets where scheduled to be sent over the communication link during the time of the trace.
 \item The {\em model size} of a DFA is its number of states, and the model size of a statechart is the sum of the model sizes of its DFAs.
\end{itemize}
The statechart introduces only very modest memory footprint. Let $M_s$ denotes the model size of a statechart, $N_{dsym}$ denotes the number of distinct symbols that were learned during the learning stage, and $N_{dfas}$ denotes the number of DFAs in the statechart. For each state the statechart needs to keep:
\begin{itemize}
 \item The estimated time to the next state (8 Bytes).
 \item For each distinct symbol: the next state and the corresponding event (8 Bytes).
 \end{itemize}
 For each DFA the statechart needs to keep its current state and the timestamp of the packet / symbol instance that led to that state (12 Bytes).

 The memory size needed for keeping the model $Model_{mem}$ is thus bounded by: $Model_{mem} = M_s \cdot (N_{dsym}+1) \cdot 8 + N_{dfas} \cdot 12$.


\begin{figure}[t]
\centering
   \includegraphics [ width=0.95\textwidth, natwidth=500,natheight=60, trim = 30 300 0 260]
    {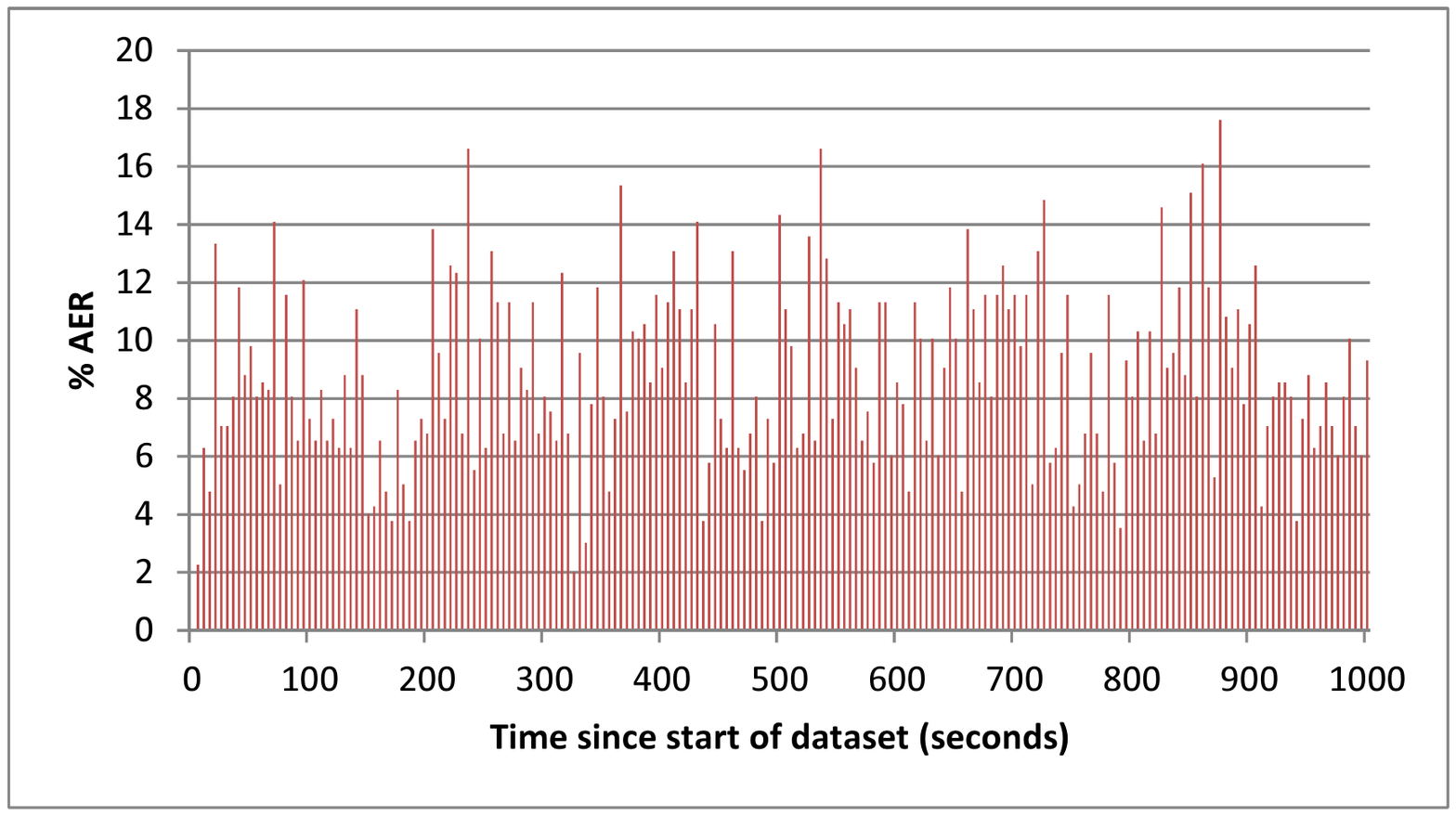}
   \hfill
\caption{Results after applying the Naive DFA model}
    \label{figure:naive-dfa-model}
   \includegraphics [ width=0.95\textwidth, natwidth=500,natheight=60, trim = 30 300 0 260]
    {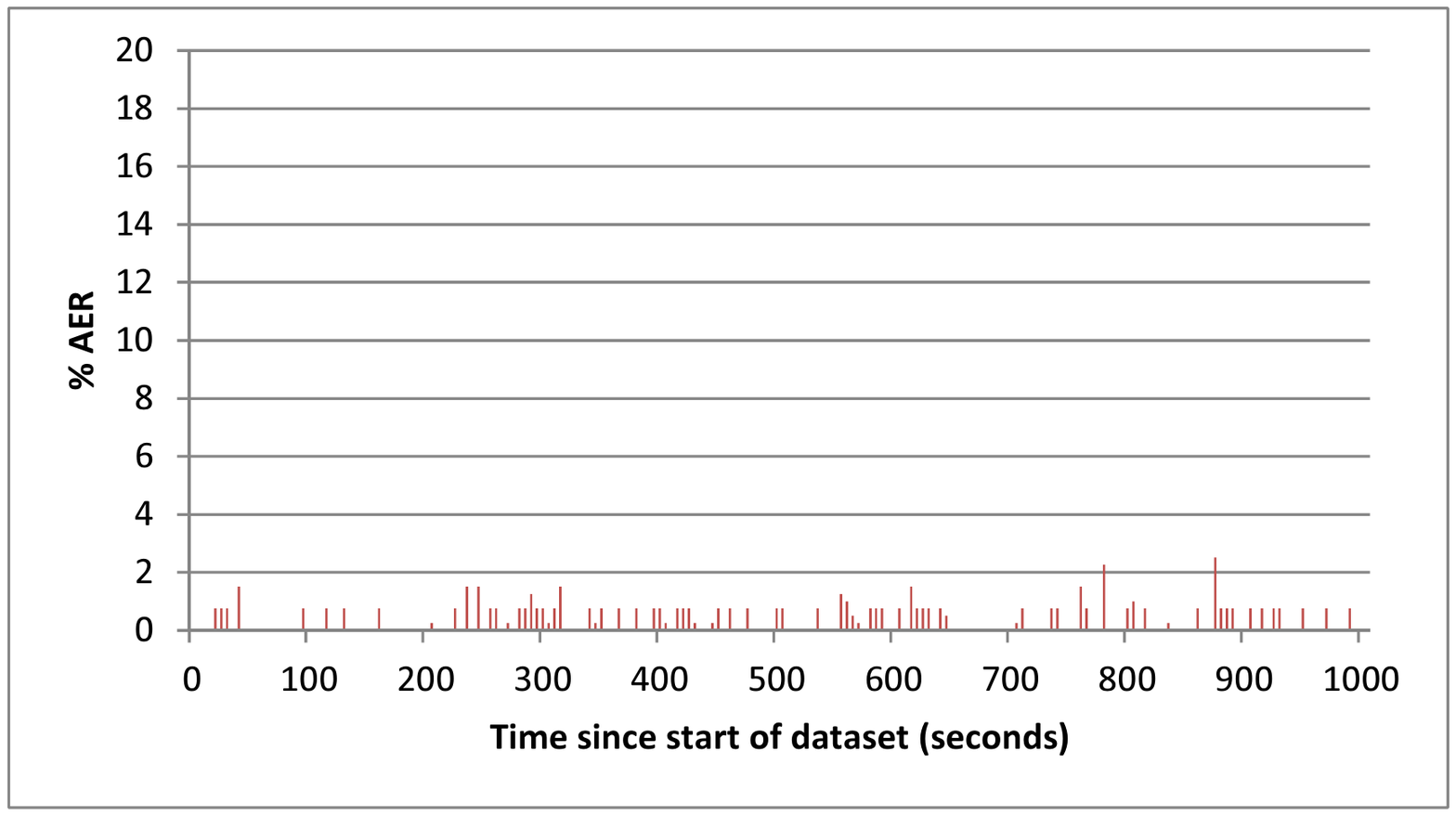}
   \hfill
\caption{Results after applying the Statechart model}
    \label{figure:synthetic-results}
\end{figure}

\begin{figure}[t]
  \centering
  \includegraphics[width=0.95\textwidth, natwidth=500,natheight=60, trim = 50 285 0 300]{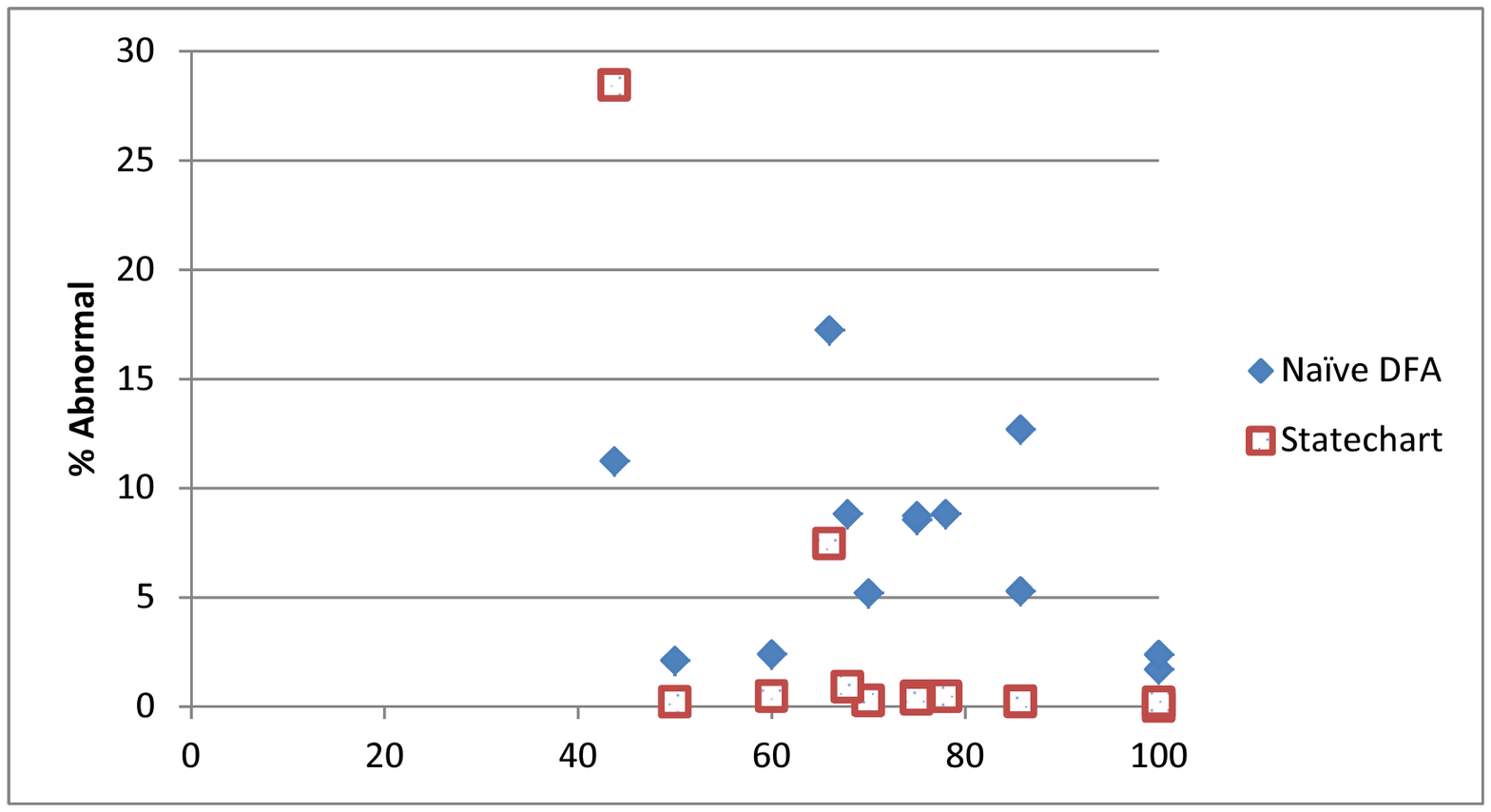}
  \hfill
  \caption{The false alarm rates as a function of the Symbol Uniqueness over the synthetic datasets.}
  \label{figure:uniqueness}

  \includegraphics[width=0.95\textwidth, natwidth=500,natheight=60, trim = 50 285 0 300]{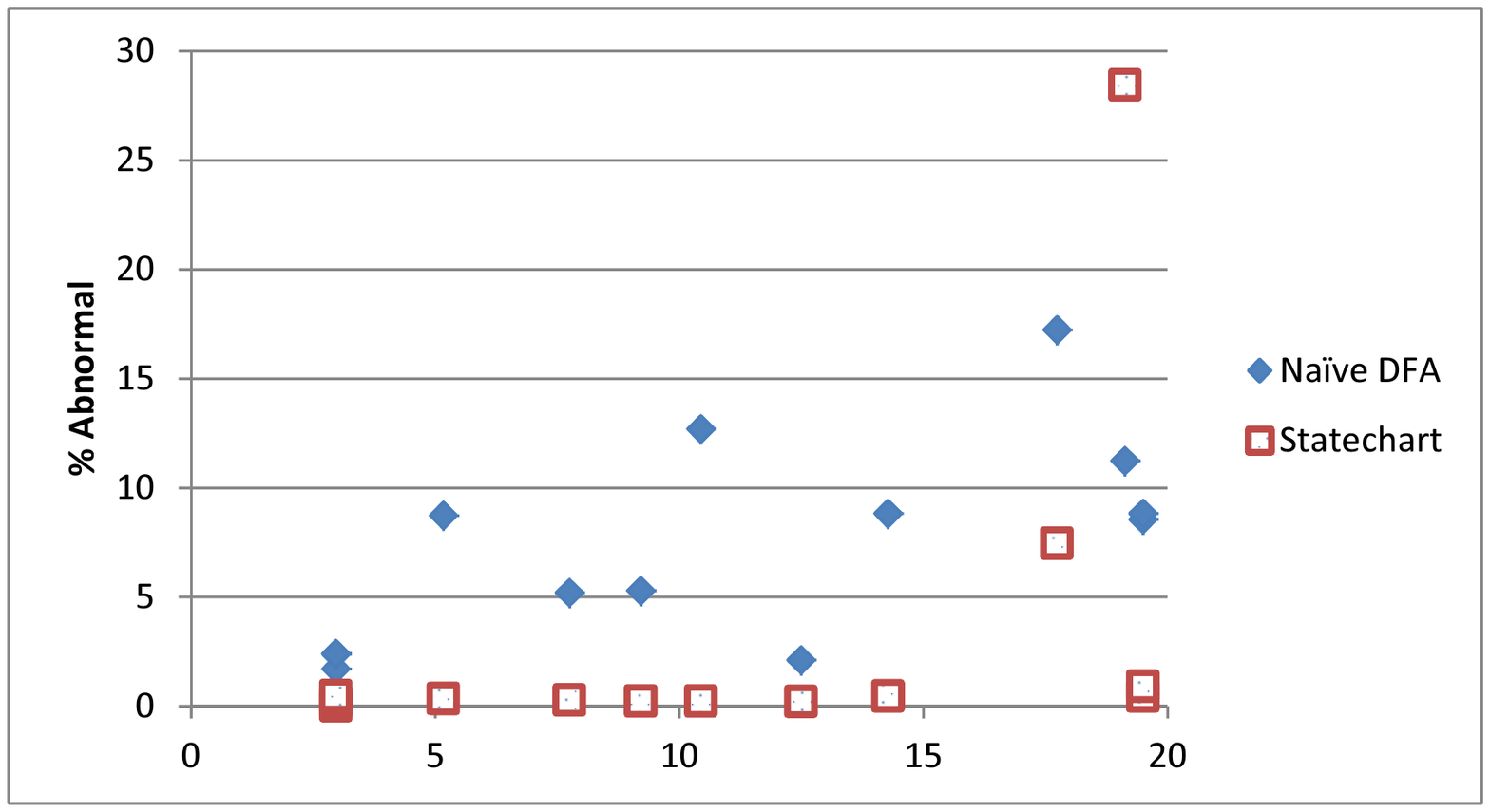}
  \caption{The false alarm rates as a function of the Time Overlap over the synthetic datasets.}
  \label{figure:collisions}
\end{figure}
\subsection{Experiments with the Synthetic Data}\label{SyntheticData}
We started our evaluation by running the DFA described by \cite{Goldenberg201363}, which we henceforth call the ``naive-DFA''. We ran the model's learning stage on the synthetic datasets with a maximum pattern length of 100 symbols and a validation window of 400 ($100 \cdot 4$) symbols. Then we ran the enforcement stage on the full datasets using the learned patterns.

When we applied the naive DFA model on the synthetic datasets it learned model sizes that are on average 3.5 times longer than the statechart model sizes for the same traces. Moreover, the {\em Statechart} model produced a much lower false-alarm rate on the same datasets. E.g, Figure \ref{figure:naive-dfa-model} and Figure \ref{figure:synthetic-results}
illustrate the results of applying the two models on synthetic dataset \#11. Each time frame on the X axis represents 5 seconds. The Y axis shows the false alarm frequency as a percentage of the Average Event Rate (AER) for each time period.

Figure \ref{figure:uniqueness} and Figure \ref{figure:collisions}
show that the {\em Statechart} managed to model the benign traffic successfully with very low false-alarm rate: up to 0.9\% in nearly all our intentionally complex scenarios. The two exception cases are of datasets \#10 (the worst result) and \#13 (second worst result) that have very low symbol uniqueness (44\% and 67\% respectively, compared to an average of 77\% for the successful cases) and a high time overlap (19.13\% and 17.74\% respectively, approximately twice the average of the successful cases of 9.76\%). In other words, only when around half of the symbols are not unique to a single pattern, {\em and\/} there is significant time overlap between patterns, does the {\em Statechart} model's performance deteriorate. In the more realistic scenarios, when symbol uniqueness is high or when the time overlap is low, the model performs extremely well.

However, this good performance was achieved when the patterns of all the multiplexed cycles were {\em known}. Hence we call this model an {\em ideal Statechart}. Next we examined scenarios where the multiplexed cycles are not known. We evaluated the performance of our Statechart model with the algorithm of sub-channel splitting. We call this model the practical Statechart model.

\begin{figure}[t]
\centering
   \includegraphics[width=0.95\textwidth, natwidth=500,natheight=60, trim = 50 300 0 260]{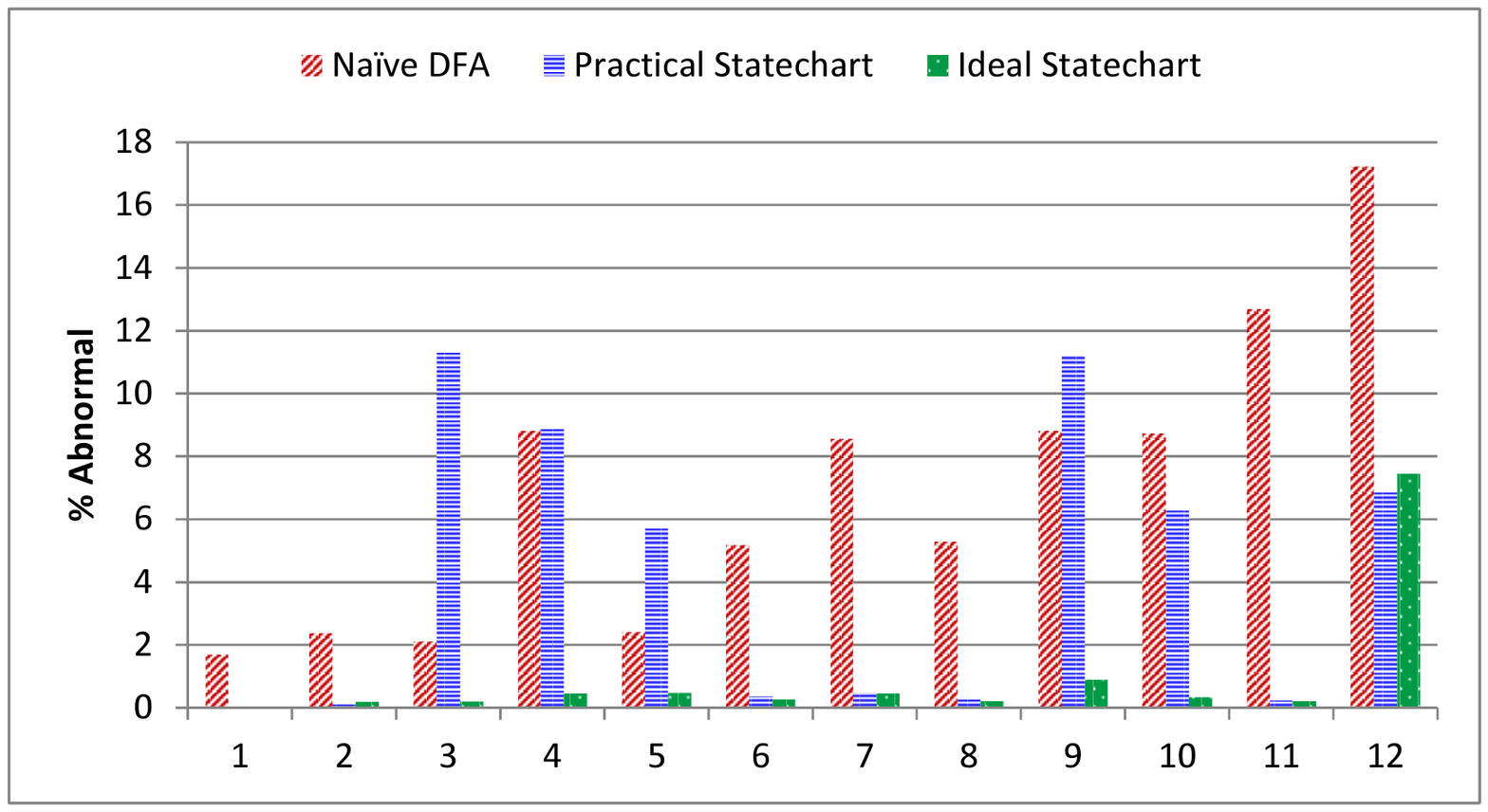}
   \hfill
\caption{Performance of the different models}
    \label{fig:compr}
\end{figure}

\subsection{Evaluating the Statechart with sub-channel splitting model}\label{practical_statechart}
 We ran the model's learning stage on the synthetic datasets with a maximum pattern length of 200 symbols and a validation window of 800 ($200 \cdot 4$) symbols. Then we ran the enforcement stage on the full datasets using the learned patterns.

When we applied the naive DFA model on the synthetic datasets it learned model sizes that are on average 3.5 times longer than the statechart model sizes for the same traces. The memory footprints of the statecharts were of sizes between 104 -- 1164 Bytes with an average size of 495 Bytes.

To test the effectiveness of our algorithms, we ran them on 800 symbols of each scenario for the learning phase, and than ran the Statechart enforcement phase on the remaining symbols of the trace. For comparison we also created the ideal Statechart using the correct cycles.

Figure \ref{fig:compr} shows comparison of the results of the three models (Naive DFA, Practical Statechart, and Ideal Statechart) of periodic traffic after applying each of the models on the SCADA traffic of each of the 12 scenarios whose characteristics are depicted in Table \ref{tab:sequence-attributes-for-synthetic-data} (scenarios \#1-\#9 and \#11-\#13, in the case of scenario \#10 the algorithm that built the Practical Statechart did not identify one of the occurrences of a specific symbol, so it could not build the Statechart for this scenario). The median percentage of detected abnormal (false alarms) on benign traffic over all the scenarios were: 5.3\%, 0.48\%, 0.28\% respectively. While the Practical Statechart model performs much better than the Naive DFA model, there is still room for improvement to achieve detection results that are closer to the impressive performance of the Ideal Statechart model.

Next we tried to evaluate what factors influence the success of the Statechart model, and the ability of our algorithms to construct it correctly.

Figure \ref{figure:uniqueness} shows the false-alarm rate of the scenarios as a function of the symbol uniqueness: the Figure shows that above 80\% symbol uniqueness the practical Statechart has excellent results - and as the symbol uniqueness drops the false alarm rate grows.

Figure \ref{figure:collisions} shows the same data points, as a function of the time overlap. The Figure shows that in general when the cycles overlap each other in time the algorithms performance degrades.

We hypothesize that in real multiplexed SCADA traces, the symbol uniqueness will be rather high (this is certainly the case for the S7 traces of \cite{KleinmannWjdfsl,KleinmannWool2015} and for the Modbus traces of \cite{Goldenberg201363}.
Hence we are optimistic that the algorithms performance on real traces will be much better than on the  synthetic traces.

\section{Conclusion}
In this paper we developed, applied, and evaluated the {\em Statechart DFA} model, which is designed specifically for anomaly detection in ICS networks. The model includes a methodology for learning individual patterns of a multiplexed cycle patterns of a ICS traffic. It uses Deterministic Time Markov Chain (DTMC) and graph theory concepts to create a DFA, or several DFA alternatives, for each cyclic pattern in a PLC-HMI channel. The result patterns are then merged to create few alternative {\em Statechart DFAs}. The alternative {\em Statechart DFAs} are then evaluated using the validation phase of the GW model, and the best {\em Statechart DFA} is selected as the Practical Statechart to be used in the enforcement phase of the ICS NIDS.

Our methodology allows automatic unsupervised learning of the individual patterns even when the number of cycles is unknown a-priori, there is no clear indication on which cycle each symbol belongs to, symbols can appear more than once in a cycle, and can appear in multiple cycles. Our experiments demonstrate that the selected Statechart model handles multiplexed ICS traffic patterns very well.

The {\em Statechart DFA} model has three promising characteristics. First, it exhibits very low false positive rates despite its high sensitivity. Second, it is extremely efficient: it has a compact representation, it keeps minimal state during the enforcement phase, and can easily work at line-speed for real-time anomaly detection.
Third, its inherent modular architecture makes it scalable for protecting highly multiplexed ICS streams.

The methodology for the automatic construction of the model, the characteristic of the model, and the validation of the model by our experiments suggest that this model can be very useful for anomaly detection in ICS networks.

We still need to test the algorithms' performance on real traces - and also test the Statechart's ability to detect true attacks. Last but not least, we need to find ways to further reduce the false positive rate.

\bibliographystyle{plain}
\bibliography{amitbib}

\begin{thebibliography}{10}

\bibitem{AfconPulse}
{Afcon Technologies}.
\newblock Pulse {HMI} software, 2015.
\newblock [Online; accessed 24-Nov-2015].

\bibitem{alcaraz_cazorla_fernandez}
Cristina Alcaraz, Lorena Cazorla, and Gerardo Fernandez.
\newblock Context-awareness using anomaly-based detectors for smart grid
  domains.
\newblock In {\em 9th International Conference on Risks and Security of
  Internet and Systems}, volume 8924, pages 17--34, Trento, 04/2015 2015.
  Springer International Publishing, Springer International Publishing.

\bibitem{atassi}
A.~Atassi, I.~H. Elhajj, A.~Chehab, and A.~Kayssi.
\newblock {\em The State of the Art in Intrusion Prevention and Detection,
  Auerbach Publications}, chapter 9: Intrusion Detection for SCADA Systems,
  pages 211--230.
\newblock Auerbach Publications, January 2014.

\bibitem{briesemeister2010detection}
L.~Briesemeister, S.~Cheung, U.~Lindqvist, and A.~Valdes.
\newblock Detection, correlation, and visualization of attacks against critical
  infrastructure systems.
\newblock In {\em 8th International Conference on Privacy Security and Trust
  (PST)}, pages 17--19, 2010.

\bibitem{byres2004use}
Eric~J Byres, Matthew Franz, and Darrin Miller.
\newblock The use of attack trees in assessing vulnerabilities in {SCADA}
  systems.
\newblock In {\em Proceedings of the International Infrastructure Survivability
  Workshop}, 2004.

\bibitem{Caselli}
M.~Caselli, E.~Zambon, and F.~Kargl.
\newblock Sequence-aware intrusion detection in industrial control systems.
\newblock In {\em Proceedings of the 1st ACM Workshop on Cyber-Physical System
  Security}, pages 13--24, New York, NY, USA, 2015.

\bibitem{chen}
Chia-Mei Chen, Han-Wei Hsiao, Peng-Yu Yang, and Ya-Hui Ou.
\newblock Defending malicious attacks in cyber physical systems.
\newblock In {\em IEEE 1st International Conference on Cyber-Physical Systems,
  Networks, and Applications (CPSNA), 2013}, pages 13--18, Aug 2013.

\bibitem{cheung2007using}
S.~Cheung, B.~Dutertre, M.~Fong, U.~Lindqvist, K.~Skinner, and A.~Valdes.
\newblock Using model-based intrusion detection for {SCADA} networks.
\newblock In {\em Proceedings of the {SCADA} Security Scientific Symposium},
  pages 127--134, 2007.

\bibitem{Dolev:1981:SPK:891726}
Danny Dolev and Andrew~C. Yao.
\newblock On the security of public key protocols.
\newblock Technical report, Stanford, CA, USA, 1981.

\bibitem{plcmarket}
Electrical engineering Blog.
\newblock The top most used {PLC} systems around the world.
\newblock Electrical installation \& energy efficiency, May 2013.
\newblock Available at:
  \url{http://engineering.electrical-equipment.org/electrical-distribution/the-top-most-used-plc-systems-around-the-world.html}.

\bibitem{Erez:2015:CVC:2822917.2823033}
Noam Erez and Avishai Wool.
\newblock Control variable classification, modeling and anomaly detection in
  modbus/tcp scada systems.
\newblock {\em Int. J. Crit. Infrastruct. Prot.}, 10(C):59--70, September 2015.

\bibitem{falliere2011w32}
N.~Falliere, L.O. Murchu, and E.~Chien.
\newblock W32. stuxnet dossier.
\newblock {\em White paper, Symantec Corp., Security Response}, 2011.

\bibitem{fovino2010modbus}
I.N. Fovino, A.~Carcano, T.~{De Lacheze Murel}, A.~Trombetta, and M.~Masera.
\newblock {Modbus/DNP3} state-based intrusion detection system.
\newblock In {\em 24th IEEE International Conference on Advanced Information
  Networking and Applications (AINA)}, pages 729--736. Ieee, 2010.

\bibitem{Goldenberg201363}
Niv Goldenberg and Avishai Wool.
\newblock Accurate modeling of modbus/tcp for intrusion detection in \{SCADA\}
  systems.
\newblock {\em International Journal of Critical Infrastructure Protection},
  6(2):63 -- 75, 2013.

\bibitem{hadziosmanovic2011}
D.~{Hadziosmanovic}, D.~{Bolzoni}, P.~H. {Hartel}, and S.~{Etalle}.
\newblock {MELISSA}: Towards automated detection of undesirable user actions in
  critical infrastructures.
\newblock In {\em Proceedings of the European Conference on Computer Network
  Defense, EC2ND 2011, Gothenburg, Sweden}, pages 41--48, USA, September 2011.
  IEEE Computer Society.

\bibitem{Harel1987}
David Harel.
\newblock Statecharts: A visual formalism for complex systems.
\newblock {\em Sci. Comput. Program.}, 8(3):231--274, June 1987.

\bibitem{hierholzer1873moglichkeit}
Carl Hierholzer and Chr Wiener.
\newblock {\"U}ber die m{\"o}glichkeit, einen linienzug ohne wiederholung und
  ohne unterbrechung zu umfahren.
\newblock {\em Mathematische Annalen}, 6(1):30--32, 1873.

\bibitem{KleinmannWjdfsl}
Amit Kleinmann and Avishai Wool.
\newblock Accurate modeling of the siemens {S7} {SCADA} protocol for intrusion
  detection and digital forensic.
\newblock {\em {JDFSL}}, 9(2):37--50, 2014.

\bibitem{KleinmannWool2015}
Amit Kleinmann and Avishai Wool.
\newblock A statechart-based anomaly detection model for multi-threaded scada
  systems.
\newblock In {\em Pre-Proceedings of the 10th International Conference on
  Critical Information Infrastructures Security (CRITIS 2015)}, pages 139--150.
  Fraunhofer IAIS, Oct 2015.

\bibitem{langner2011stuxnet}
Ralph Langner.
\newblock Stuxnet: Dissecting a cyberwarfare weapon.
\newblock {\em Security \& Privacy, IEEE}, 9(3):49--51, 2011.

\bibitem{SicherheitDeutschland}
T.~D. Maiziere.
\newblock Die lage der it-sicherheit in deutschland 2014.
\newblock Technical report, Bundesamt fur Sicherheit in der
  Informationstechnik, 2014.

\bibitem{RobertT}
Robert~T. Marsh.
\newblock Critical foundations: Protecting america's infrastructures - the
  report of the president's commission on critical infrastructure protection.
\newblock Technical report, October 1997.

\bibitem{mukherjee1994network}
Biswanath Mukherjee, L~Todd Heberlein, and Karl~N Levitt.
\newblock Network intrusion detection.
\newblock {\em Network, IEEE}, 8(3):26--41, 1994.

\bibitem{emerald}
Phillip~{A.} Porras and Peter~{G.} Neumann.
\newblock {EMERALD:} event monitoring enabling responses to anomalous live
  disturbances.
\newblock In {\em 1997 National Information Systems Security Conference}, oct
  1997.

\bibitem{roesch}
Martin Roesch.
\newblock Snort - lightweight intrusion detection for networks.
\newblock In {\em Proceedings of the 13th USENIX Conference on System
  Administration}, LISA '99, pages 229--238, Berkeley, CA, USA, 1999. USENIX
  Association.

\bibitem{sommer}
R.~Sommer and V.~Paxson.
\newblock Outside the closed world: On using machine learning for network
  intrusion detection.
\newblock In {\em Security and Privacy (SP), 2010 IEEE Symposium on}, pages
  305--316, May 2010.

\bibitem{stouffer}
K.~A. Stouffer, J.~A. Falco, and K.~A. Scarfone.
\newblock Guide to industrial control systems ({ICS}) security.
\newblock Technical Report 800-82, National Institute of Standards and
  Technology (NIST), Gaithersburg, MD, May 2013.

\bibitem{valdes2009communication}
A.~Valdes and S.~Cheung.
\newblock Communication pattern anomaly detection in process control systems.
\newblock In {\em IEEE Conference on Technologies for Homeland Security (HST)},
  pages 22--29. IEEE, 2009.

\bibitem{s7dissector}
T.~Wiens.
\newblock S7comm wireshark dissector plugin, January 2014.
\newblock Available at: \url{http://sourceforge.net/projects/s7commwireshark}.

\bibitem{vlq}
Wikipedia.
\newblock Variable-length quantity --- {W}ikipedia{,} the free encyclopedia,
  2015.
\newblock [Online; accessed 5-May-2015].

\bibitem{yang2006anomaly}
D.~Yang, A.~Usynin, and J.W. Hines.
\newblock Anomaly-based intrusion detection for {SCADA} systems.
\newblock In {\em 5th Intl. Topical Meeting on Nuclear Plant Instrumentation,
  Control and Human Machine Interface Technologies}, pages 12--16, 2006.

\bibitem{ye}
N.~Ye, Y.~Zhang, and C.M. Borror.
\newblock Robustness of the markov-chain model for cyber-attack detection.
\newblock {\em IEEE Transactions on Reliability}, 53(1):116--123, 2004.

\end{thebibliography}

\end{document}